\documentclass[sn-mathphys-num]{sn-jnl}

\usepackage{geometry} 
\usepackage{graphicx}%
\usepackage{amsmath,amssymb,amsfonts}%
\usepackage{amsthm}%
\usepackage{mathrsfs}%
\usepackage[title]{appendix}%
\usepackage{xcolor}%
\usepackage{dsfont}
\usepackage{xspace}
\usepackage[most]{tcolorbox}
\tcbuselibrary{breakable}

\newtcbtheorem[]{tcexample}{Example}{breakable, colback=orange!3!white,colframe=orange!85!black, fonttitle=\bfseries}{Exa}

\usepackage{tikz,xcolor}
\definecolor{lime}{HTML}{A6CE39}
\DeclareRobustCommand{\orcidicon}{%
	\begin{tikzpicture}
	\draw[lime, fill=lime] (0,0) 
	circle [radius=0.16] 
	node[white] {{\fontfamily{qag}\selectfont \tiny ID}};
	\draw[white, fill=white] (-0.0625,0.095) 
	circle [radius=0.007];
	\end{tikzpicture}
	\hspace{-2mm}
}
\foreach \x in {A, ..., Z}{%
	\expandafter\xdef\csname orcid\x\endcsname{\noexpand\href{https://orcid.org/\csname orcidauthor\x\endcsname}{\noexpand\orcidicon}}
}

\begin{document}

\title[Limits to the validity of gravitational redshift 
as a quantum-optical multimode mixer]{Limits to the validity of gravitational redshift as a quantum-optical multimode mixer}

\author[1,2]{\fnm{Nils}  \sur{Leber}\orcidE{}}

\author[3]{\fnm{Luis Adri\'an} \sur{Alan\'is Rodr\'iguez}\orcidF{}}

\author[2]{\fnm{Alessandro} \sur{ Ferreri}\orcidA{}}

\author[4]{\fnm{Andreas Wolfgang} \sur{  Schell}\orcidD{}}

\author*[2,1]{\fnm{David Edward} \sur{  Bruschi}\orcidB{}}\email{david.edward.bruschi@posteo.net}\email{d.e.bruschi@fz-juelich.de}

\affil*[1]{Theoretical Physics, Universit\"at des Saarlandes, 66123 Saarbr\"ucken, Germany}

\affil[2]{Institute for Quantum Computing Analytics (PGI-12), Forschungszentrum J\"ulich, 52425 J\"ulich, Germany}

\affil[3]{Institute for Theoretical Physics, Faculty of Mathematics and Natural Sciences, University of Cologne, Zülpicherstrasse 77, 50937 Cologne, Germany}

\affil[4]{Division of light-matter interaction, Johannes Kepler Universit\"at Linz, 4040 Linz, Austria}

\abstract{We analyze the domain of validity of a quantum optical model that describes the effects of gravitational redshift on the quantum state of photons that propagate in curved spacetime. This model assumes that the modes defining the initial state of the photon are mixed with an auxiliary environment mode via an effective multimode mixer. We find that the model, as proposed, is consistent only to first order for small redshift, where the range of validity is conditional not only to the gravitational parameters, but also to those that define the photonic modes. We identify the problem and provide a partial solution in terms of a necessary condition on the transformation matrix representing the process, which requires the use of a number of auxiliary modes that is at least equal to the number of modes that define the photonic state. We conclude by discussing implications for theoretical quantum optics and photonics in curved spacetime, as well as for the development of quantum technologies.}

\keywords{Gravitational redshift, Quantum field theory, Quantum Optics, Photonics in curved spacetimes}

\maketitle

\section*{Introduction}
The development of quantum technologies requires significant efforts that range from building the necessary theoretical methods and tools to determining the best transmission channels to be employed or the best environment for the deployment of the hardware. Foreseen technologies can be broadly grouped into two categories: those that will be solely bound to Earth, and those that will require the infrastructure to be at least partially located in space. Free-space links are the natural candidates when communication occurs between users located at large distances. In such cases, photons are the system of choice to carry the desired quantum information between distant nodes in a network. Applications based on free-space links include quantum communciation \cite{Yin:Ren:2017,Zhang:Xu:2018,Sidhu2021,Pirandola:2021,Cao:Yin:2023}, distributed quantum computing \cite{boschero2024distributedquantumcomputingapplications,Caleffi2024}, and the quantum internet \cite{Trinh:2024} to name a few. As a consequence, space-based photonics has become a key area of research within the broad effort to deploy a global quantum network \cite{Yin:Ren:2017,Simon_2017,Liao:Yong:2017,Gundogan:Sidhu:2021}. 

A key aspect of space-based science and technology is the unavoidable background curvature due to gravity. In the specific case of the space surrounding the Earth, curvature is weak but nevertheless nonvanishing \cite{Nichols:Owen:2011}. In general, quantum information tasks are proposed and studied with the implicit assumption that the coupling of the involved systems with gravity is negligible. While this assumption might be appropriate for protocols that are based either on Earth or employ short links between users, the situation can be expected to change significantly when moving part of the infrastructure to space. Hence, one important question is: \emph{What are the effects, if any, of spacetime curvature and gravity on the protocols of interest}?

\begin{figure}[t!]
    \centering
    \includegraphics[width=0.5
    \linewidth]{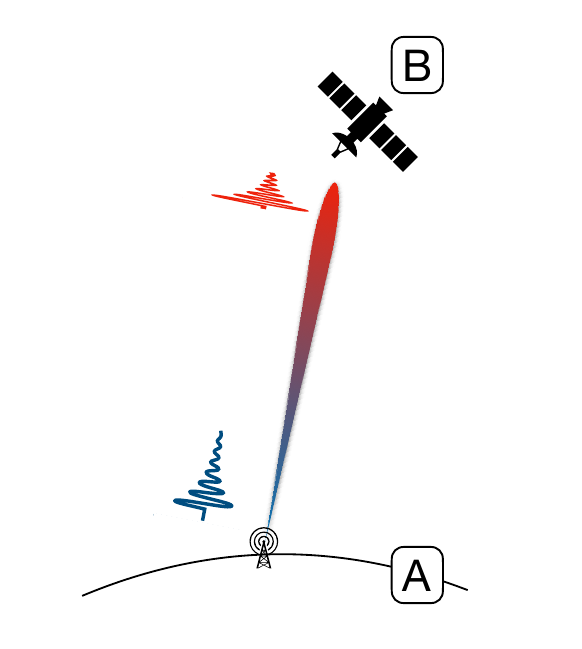}
    \caption{\textbf{General scheme}: in this work we consider photons that propagate between two points A and B in a (weak) gravitational field. Photon exchange between an Earth-based user, Alice, and a satellite-based user, Bob, is a paramount example of such general situation.}
    \label{fig:initial}
\end{figure}

In the past few years, a quantum-optical model has been put forward to understand the effects of gravitational redshift on the quantum state of photons that propagate in (weakly) curved spacetime \cite{Bruschi:Ralph:2014,Alanis:Schell:Bruschi:2023}, which we refer to as the \emph{Quantum-Optical Gravitational Redshift Model (QOGRM)}. Gravitational redshift is a celebrated prediction of general relativity, where a photon sent by a user at one location is received with a different frequency by a second user located at another point in a gravitational field \cite{Einstein:1908,Wald:1995}. The change in frequency profile, or the mode of the photon, occurs concomitantly with the change in its quantum state. The effective description of this process has been shown to correspond to a multimode mixer \cite{Bruschi:Schell:2023}, that is, a quantum optical process where two or more modes of the electromagnetic field are mixed linearly without a variation in the total number of excitations. Processes with such properties are known as a \textit{passive} transformation in the language of quantum optics \cite{Adesso:Ragy:2014}. The main idea is that the overlap of the photonic mode with the original one, as it travels in curved spacetime, changes and the new mode increasingly has support on other modes. In particular, it was suggested that it is possible to consider $N$ initial modes of choice and collect all of the remaining (infinite) modes into a single ``environment'' mode that absorbs the information leaking out of the subsystem of interest due to the mode-mixing effect. A pictorial representation of the basic photon-exchange process can be found in Figure~\ref{fig:initial}.

Subsequent work further characterized this effect at the overlap of relativity and quantum field theory. For example, it has been shown that it can be exploited for precise measurements of distances or the mass of the Earth using quantum states of light as input \cite{Bruschi:Datta:2014,Kohlrus:Bruschi:2017,Kohlrus:Bruschi:2019}, for interference experiments \cite{Barzel:Bruschi:2022}, for weakening the non-Markovian dynamical behaviour of photons \cite{PhysRevD.108.126011}, or for enhancing the precision of phase estimation of two qubit system states \cite{PhysRevD.111.026007}. Regardless of the progress \cite{Alanis:Schell:Bruschi:2023,molaei2024photongravitycouplingschwarzschildspacetime}, an underlying unchallenged assumption has always been implicitly made: the multimode-mixing model applies regardless of the magnitude of the redshift or of other relevant parameters of the problem, such as the bandwidth, average frequency of the spectrum, or the particular shape of the spectrum itself. 

In this work we show that gravitational redshift cannot be modelled as a multimode mixer for all values of the gravitational redshift if only one environment mode is considered additionally to the $N$ modes 
of interest, a setup called the \emph{$N+1$-decomposition} of the channel. This is a consequence of the ``rigidity'' of the transformation of interest: in fact, QOGRM posits that each mode of the electromagnetic field transforms \textit{in the same fashion}, that is, all frequencies are consistently either redshifted or blueshifted, and therefore any wave-packet of light undergoes the same type of transformation. We show that this assumption, while natural in light of our current understanding of the standard gravitational redshift effect as predicted by general relativity, leads to a transformation that becomes non-unitary for values of the redshift that are large enough.  Interestingly, the transition beyond the validity and invalidity of QOGRM strongly depends on the interplay between the magnitude of the redshift and the parameters that determine the wave-packet itself.

We provide a partial solution to the problem by extending the $N+1$-decomposition to an $N+M$-decomposition of the channel, and we show that this extension provides a necessary solution to recovering unitarity for all values of the redshift when $M\geq N$. In other words, unitarity of the whole transformation requires that there is at least the same amount of environment modes as to those of interest, but is not guaranteed by this property alone. This result increases the complexity of designing realistic quantum communication protocols or other quantum information protocols, but the price to pay is not significant: in the case of few modes of interest it can be possible to derive the environment modes analytically by applying the constraints that define the unitarity of the transformation. When the number of input modes increases, on the other hand, numerical methods have to be employed. In both cases, the nature of the environment modes remains undetermined.
We finally discuss our results and their implications not only for the theoretical part of the problem at hand, but also for the development of future quantum technologies.

This work is organized as follows. In Section~\ref{sec previous material} we present the mathematical formalism required to model the gravitational redshift induced on photons by propagation in curved spacetime. We further characterize this effect in Section~\ref{sec characterizing redshift} by providing an analysis of the frequency change and the perturbative expansion of the overlap. We then devote Section~\ref{sec new work} to the discussion of the domain of validity of the aforementioned transformation. Section~\ref{section:solution} proposes a partial solution to the problem. Additional considerations and an outlook to this work are given in Section~\ref{considerations}. Lastly, we collect our concluding remarks in Section~\ref{sec conclusions}.

\section{Background material}\label{sec previous material}
We discuss the tools used in this work. A detailed introduction can be found in the literature \cite{Birrell:Davies:1982,Wald:1995,Alanis:Schell:Bruschi:2023}.\footnote{In this work we set $\hbar=c=1$. We employ Einstein's summation convention. The metric has signature $(-,+,+,+)$.}

\subsection{Modelling photons in weakly curved spacetime}
Photons, at the fundamental level, are excitations of a quantum field. A photon in configuration space can be understood as a wavepacket of sharp frequency modes with a given helicity that propagates at the speed of light. A proper treatment of the subject would require the use of spin-$1$ fields, which would significantly increase the complexity of the computations without necessarily adding new qualitatively different insights \cite{Smith:Raymer:2007,Milburn:2008}. In this regard we note that, in the context of General Relativity, massless scalar fields are often employed to provide a qualitative analysis of the phenomena of interest and without loss of generality \cite{Misner:Thorne:1973}. Therefore, for the purposes of our work and for the sake of simplicity, we follow suit and model photons as excitations of a real massless scalar field.

We consider a real, massless, and scalar quantum field $\Phi(x^\mu)$ defined in a $3+1$ dimensional spacetime with coordinates $x^\mu$ and metric $\boldsymbol{g}=(g_{\mu\nu}(x^\sigma))$. The free field equations read $\square\Phi(x^\mu)=0$, where $\square:=(-g)^{-1/2}\partial_\mu (-g)^{1/2} g^{\mu\nu}\partial_\nu$ and $g:=\text{det}(\boldsymbol{g})$, which can be further elaborated by employing a basis $\{\phi_{\boldsymbol{k}}(x^\mu)\}$ of mode solutions each of which satisfies $\square\phi_{\boldsymbol{k}}(x^\mu)=0$. Note here that $\boldsymbol{k}$ is a collection of (continuous and/or discrete) indices that label our modes. We employ the covariant inner product $(f_1,f_2)=-i\int_V \textrm{d}V^\mu \sqrt{-g_V} \, (f_1\partial_\mu f_2^*-(\partial_\mu f_1) f_2^*)$ between two functions $f_1,f_2$ defined over the manifold, where $V$ is a Cauchy hypersurface of integration, $\textrm{d}V^\mu:=\textrm{d}V n^\mu$ is the integration element with $n^\mu$ a future-pointing, unit-norm, normal vector to $V$, and $g_V$ is the determinant of the induced metric on $V$. This allows us to obtain the mode orthonormalization conditions $(\phi_{\boldsymbol{k}},\phi_{\boldsymbol{k}'})=(\phi_{\boldsymbol{k}}^*,\phi_{\boldsymbol{k}'}^*)=0$, while $(\phi_{\boldsymbol{k}},\phi_{\boldsymbol{k}'}^*)=\delta(\boldsymbol{k}-\boldsymbol{k}')$. Here, the nature of the delta-normalization depends on the labels $\boldsymbol{k}$ being continuous and/or discrete.
We can then meaningfully decompose the quantum field as
\begin{align}
    \hat{\Phi}(x^\mu)=\int_\Sigma \textrm{d}\Sigma\left[\phi_{\boldsymbol{k}}(x^\mu)\hat{a}_{\boldsymbol{k}}+\phi^*_{\boldsymbol{k}}(x^\mu)\hat{a}_{\boldsymbol{k}}^\dag\right],
\end{align}
where $\Sigma$ is an appropriate $3$-dimensional integration hypersurface in Fourier space parameterized by coordinates $\boldsymbol{k}$.

The annihilation and creation operators $\hat{a}_{\boldsymbol{k}},\hat{a}_{\boldsymbol{k}}^\dag$ satisfy the canonical commutation relations $[\hat{a}_{\boldsymbol{k}},\hat{a}_{\boldsymbol{k}'}^\dag]=\delta(\boldsymbol{k}-\boldsymbol{k}')$, where all others vanish. This tempts us to define the vacuum state $|0\rangle$ through the constraint $\hat{a}_{\boldsymbol{k}}|0\rangle=0$ for all $\boldsymbol{k}$. Unfortunately, there is no natural choice of the vacuum state in arbitrary curved spacetime \cite{Birrell:Davies:1982}. Among the consequences of the non uniqueness of the vacuum one has, for example, the potential disagreement of different observers on the particle content of the state of the system, including the vacuum state itself \cite{Birrell:Davies:1982}. Here we focus on scenarios that allow for a description that can be handled analytically, and where particle creation due to gravitational redshift is negligible or nonexistent altogether. These scenarios include weakly curved spacetimes, such as those around a massive planet \cite{Misner:Thorne:1973}, or regions that are static:
\begin{itemize}
    \item \textbf{Weak curvature}. One scenario of interest is that of \textit{linearized gravity}, which is commonly employed in many areas of modern physics, for example when considering gravitational waves \cite{Flanagan:Hughes:2005} or linearized quantum gravity \cite{guptaQuantizationEinsteinGravitational1952}. This regime well adpats to regions of spacetime that are weakly curved, thus restricting the kinematics to a scenario where the background metric has the expression $g_{\mu\nu}=\eta_{\mu\nu}+\epsilon h_{\mu\nu}$, $\epsilon\ll1$ is a small control parameter, and $h_{\mu\nu}$ is the linearized contribution to the metric.
    \item \textbf{Local timelike Killing vector}. A Killing vector $K:=\partial_\xi$ is vector field along which ``the metric does not change''. Abstractly one has $\mathcal{L}_K\boldsymbol{g}=0$, which is equivalent to $K_{\mu,\nu}+K_{\nu,\mu}=0$, \cite{Wald:1995}. A timelike Killing vector field $K=\partial_\xi$ allows for a preferred or natural notion of time, parametrized by $\xi$, which is of crucial importance when quantizing a system in curved spacetime. In such case, it is convenient to choose the orthonormalized mode functions $\phi_{\boldsymbol{k}}(x^\mu)$ as solutions to the field equations, which satisfy the eigenvalue constraint $i\partial_\xi\phi_{\boldsymbol{k}}(x^\mu)=\omega_{\boldsymbol{k}} \phi_{\boldsymbol{k}}(x^\mu)$, where $\omega_{\boldsymbol{k}}$ are positive eigenvalues. When the spacetime is static, one also has that $K$ can be hypersurface-orthogonal and therefore foliate the spacetime into hypersurfaces $V$ of constant ``time'' $\xi$. The variables $\boldsymbol{k}\equiv(k_1,k_2,k_3)$ parametrize the $3$-dimensional Fourier space. Recall that in the simple case of flat spacetime, for example, one has $\omega_{\boldsymbol{k}}=|\boldsymbol{k}|$ for the frequency $\omega_{\boldsymbol{k}}$ associated to the field modes. If spacetime is weakly curved, it might not have a Killing vector. However, in this case one can choose to work with mode solutions that are a obtained as a perturbation of the flat-spacetime plane wave solutions $\phi_{\boldsymbol{k}}(x^\mu)\propto\exp[ik_\mu x^\mu]$, which satisfy $i\partial_t\phi_{\boldsymbol{k}}(x^\mu)=\omega_{\boldsymbol{k}}\phi_{\boldsymbol{k}}(x^\mu)$. Note that this second case does not necessarily imply weak curvature.
\end{itemize}
We have laid out the framework within which we will be working. We thus have field modes $\phi_{\boldsymbol{k}}(x^\mu)$ that are solutions to the field equations and are eigenfunctions of a timelike Killing vector $K$ \emph{at least in the region of spacetime of interest}. We assume that it is therefore possible to identify a preferred vacuum state $|0\rangle$ given the constraints just laid out (e.g., in the case of linearized gravity we choose the Minkowski vacuum) and define the single-particle states via the relation $|1_{\boldsymbol{k}}\rangle:=\hat{a}_{\boldsymbol{k}}^\dag|0\rangle$. Such sharp-momentum single-particle states are not normalized since $\langle1_{\boldsymbol{k}}|1_{\boldsymbol{k}'}\rangle=\delta^3(\boldsymbol{k}-\boldsymbol{k}')$. These states can be useful for proof-of-principle considerations regarding quantum field theory in curved spacetime, however, they are not useful when attempting to analyze physical processes through the lens of quantum information. 

We therefore introduce a well-defined notion of photon as follows. A photon is a broad-band excitation of the field determined by a spectrum $F(\boldsymbol{k})$ at a chosen time (here we always assume that photons are initially defined at $t=0$) initially localized around a position $\boldsymbol{x}_0$, and is associated to the bosonic operator
\begin{align}
    \hat{A}_F(\boldsymbol{x}_0):=\int \textrm{d}^3k\, F(\boldsymbol{k}) \hat{a}_{\boldsymbol{k}},
\end{align}
where we require $\langle F,F\rangle\equiv\int \textrm{d}^3k |F(\boldsymbol{k})|^2=1$ in order to have $[\hat{A}_F,\hat{A}_F^\dag]=1$. This latter aspect is crucial as discussed below. Here have introduced the notation $\langle F,F'\rangle:=\int \textrm{d}^3k F^*(\boldsymbol{k}) F'(\boldsymbol{k})$ for later convenience. With these constraints, one can interpret $\rho(\boldsymbol{k}):=|F(\boldsymbol{k})|^2/(2\pi)^3$ as the distribution of momenta that characterizes a photon, and therefore $\tilde{F}(\boldsymbol{x}):=(2\pi)^{-3}\int \textrm{d}^3k F(\boldsymbol{k})e^{-i \boldsymbol{k}\cdot\boldsymbol{x}}$ is the mode function in configuration space, where $\boldsymbol{x}$ are the coordinates on the hypersurface $V$. This implies that $\langle \tilde{F},\tilde{F}\rangle=\int_V \textrm{d}^3x |\tilde{F}(\boldsymbol{x})|^2=1$, which means that we can interpret $\tilde{\rho}(\boldsymbol{x}):=|\tilde{F}(\boldsymbol{x})|^2$ as the probability density function of finding the photon at position $\boldsymbol{x}$ at $t=0$.
One particle states at time $t=0$ of a photon in mode $F$ are therefore constructed in the usual fashion as $|1_F\rangle\equiv\hat{A}_F^\dag|0\rangle$.

Photons are $3$-dimensional particles, in the sense that the spatial mode distribution $\tilde{F}(\boldsymbol{x})$ extends, in principle, in three spatial dimensions. While a full $3$-dimensional analysis in arbitrary curved spacetimes would allow for realistic characterization of the kinematics of photons that propagate, we make the following further assumption, which allows us to obtain explicit results:
\begin{itemize}
    \item \textbf{One-dimensional localization}. We assume that the photons under consideration are   localized mostly along the direction of propagation. That is, the typical spatial extension $\ell_\perp$ of the photon in the directions perpendicular to that of propagation is much smaller than the length $\ell$ of the photon itself along the direction of propagation, i.e., $\ell_\perp/\ell\ll1$. 
\end{itemize}
Given the assumptions above, we work in Minkowski coordinates $(t,\boldsymbol{x})$ 
, where $t$ is now the usual time variable and $P=\frac{d}{d\lambda}$ is the four momentum of the photon, and $\lambda$ is a chosen affine parameter along its path \cite{Wald:1995}. We find it convenient to introduce $P_\perp:=\frac{d}{d\lambda_\perp}$ as the projection of $P$ on the hypersurface $V$, i.e., the natural spatial hypersurfaces of all observers measuring time $t$. It has been shown that, in such a regime, the photon operator can be effectively written at an arbitrary time $t$ as
\begin{align}\label{time:dependent:extended:photon:operators}
    \hat{A}_F(t,\lambda_\perp) := \int_0^{+\infty} d\omega\, F(\omega)\,e^{-i(t-\lambda_\perp)\omega}\,\hat{a}_{\omega},
\end{align}
where $\omega$ is the frequency of the photonic contributions $\hat{a}_\omega$, and we have $[\hat{a}_\omega,\hat{a}_{\omega'}^\dag]=\delta(\omega-\omega')$ while all other commutators vanish \cite{Alanis:Schell:Bruschi:2023}. Here, the variables $\lambda_\perp$ and $t$ identify the position of the photon along the trajectory via the constraint $t-\lambda_\perp=-x_0$, where $x_0$ is a constant.\footnote{Note that $u:=t-\lambda_\perp$ is an \textit{outgoing null coordinate} that normally is used for cases such as those studied here. In our specific work this change of coordinates is not needed since we do not manipulate these expressions further.} It should be clear from our discussion above that pointlike (or very small, in our case) photons propagate along paths of constant null coordinate $u$ \cite{Wald:1995}.
The photon operators \eqref{time:dependent:extended:photon:operators} enable us to compute the equal-time and equal-position commutators
\begin{align*}
    \left[\hat{A}_{F}(t,\lambda_\perp),\hat{A}^\dag_{F'}(t,\lambda_\perp)\right]=\langle F,F'\rangle
\end{align*}
while all others vanish. Clearly, when $F,F'$ are orthogonal we have $\langle F,F'\rangle=0$ and therefore the commutator vanishes. Notice that the requirement for equal time and position is an idealization necessary for the expressions to be manageable analytically. 

We now note that the Hilbert space $\mathcal{H}$ corresponding to the scalar (photonic) field in question is infinite dimensional. We are free to choose any orthonormal complete basis $\mathcal{B}$ for the field, and in general we select a set of functions $\{F_{\vec{\zeta}}\}$ parametrized by the vector of indices $\vec{\zeta}$ that satisfies $\langle F_{\vec{\zeta}}, F_{\vec{\zeta}'}\rangle = \delta_{\vec{\zeta},\vec{\zeta}'}$.\footnote{Once more, the nature of the delta $\delta_{\vec{\zeta},\vec{\zeta}'}$ depends on the nature of the indices, and we give it for understood that a more rigorous notation should be used when such aspects have been determined by the specific problem at hand.} This implies that $\bigl[\hat{A}_{\vec{\zeta}}(t,\lambda_\perp),\hat{A}^\dag_{\vec{\zeta}'}(t,\lambda_\perp)\bigr]= \delta_{\vec{\zeta},\vec{\zeta}'}$, thus allowing us to 
use the operators $\hat{A}_{\vec{\zeta}}(t,\lambda_\perp)\equiv\hat{A}_{F_{\vec{\zeta}}}(t,\lambda_\perp)$ to define the vacuum $|0\rangle$ as measured locally in space at each point in time $t$ via the constraint $\hat{A}_{\vec{\zeta}}(t,\lambda_\perp)|0\rangle=0$ and to use the operators $\hat{A}^\dag_{\vec{\zeta}}(t,\lambda_\perp)$ to create a particle at time $t$ and location $\lambda_\perp$  
via the expression
\begin{align*}
|1_{\vec{\zeta}}(t)\rangle\equiv\hat{A}^\dag_{\vec{\zeta}}(t,\lambda_\perp)|0\rangle.  
\end{align*}
Multi-photon states are then created using the operators $\hat{A}^\dag_{\vec{\zeta}}(t,\lambda_\perp)$ within the standard Fock-space construction~\cite{Birrell:Davies:1982}.

\subsection{Modelling gravitational redshift of quantum photons}
We are now interested in modelling the effects of gravitational redshift on the quantum state of the photon. Gravitational redshift is an effect predicted by classical general relativity \cite{Einstein:1908,Wald:1995} that has been now firmly established by precision measurements \cite{bothwellResolvingGravitationalRedshift2022,PhysRevLett.45.2081,zhengLabbasedTestGravitational2023, mullerPrecisionMeasurementGravitational2010}. At its core, this effect can be seen arising as a consequence of the mismatch in local time-keeping of two \emph{pointlike} observers located at different heights in the gravitational potential \cite{Wald:1995}. The relation between local frequencies $\Omega_\text{A}$ and $\Omega_\text{B}$, as measured by two different pointlike observers Alice and Bob respectively, is given by
\begin{equation}\label{gravitational:redshift}
    \chi^2:= \frac{\Omega_\text{B}}{\Omega_\text{A}} = \frac{P_\mu U^\mu_{\text{B}}|_{x^\mu_\textrm{f}}}{P_\mu U^\mu_{\text{A}}|_{x^\mu_\textrm{i}}},
\end{equation}
where $\chi^2$  is the redshift factor. Here, $U_K$ is the four velocity of Alice ($K=\textrm{A}$) and Bob ($K=\textrm{B}$), while $P=P^\mu\partial_\mu$ is the four-momentum of the photon, and the photon is emitted at position $x^\mu_\textrm{i}$ and detected at position $x^\mu_\textrm{f}$. Note that $\chi^2$ can be re-written as $\chi^2=1+z$, where $z$ is the also called the redshift in a significant part of the literature \cite{Wald:1995,Will_2014,DelvaSchoenemannDilssner_2018}.\footnote{Our convention for $z$ implies that $z>0$ for blueshift processes, while $z<0$ for redshift processes, which is opposite to the conventions typically found in the literature.} 
We choose to retain the notation $\chi^2$ for the redshift in agreement with recent literature on which this work is based \cite{Bruschi:Ralph:2014,Alanis:Schell:Bruschi:2023}.

We now focus on the effects of redshift on the quantum state of the photon. As mentioned above, we have constructed the operators $\hat{A}_F^\dag$ to create local photon excitations with mode function $F$ by acting on the vacuum state $|0\rangle$. Let a photon propagate between two users Alice and Bob located at points A and B in curved spacetime. In general, Alice and Bob have agreed on a photon to be shared whose particular frequency profile $F$ is determined before the process takes place. Therefore, Bob can compare the received photon $|1_{F'}\rangle$ with the expected photon $|1_F\rangle$ and determine if a change has occurred. It has been shown that the transformation $T(\chi): \hat{A}_F \rightarrow \hat{A}_{F'}$ that models the action of the redshift can be implemented by a unitary operator $\hat{U}(\chi)$ via the relation 
\begin{align}\label{eq frequency change}
    \hat{A}_{F'} := \hat{U}^\dagger (\chi) \, \hat{A}_F\, \hat{U}(\chi) \equiv \int_0^{+\infty} d\omega\, F'(\omega)\, \hat{a}_\omega,
\end{align}
where the transformation
\begin{align}\label{fundamental:relation}
    F'(\omega)= \chi^{-1} F(\chi^{-2}\omega)
\end{align}
has been obtained as the constitutive relation of this process. Note that we ignore the propagation phase $e^{-i(t-\lambda_\perp)\omega}$ since this transforms covariantly with the redshift and is irrelevant to photon comparison~\cite{Alanis:Schell:Bruschi:2023}.

\begin{figure}[t!]
    \centering
    \includegraphics[width=0.8\linewidth]{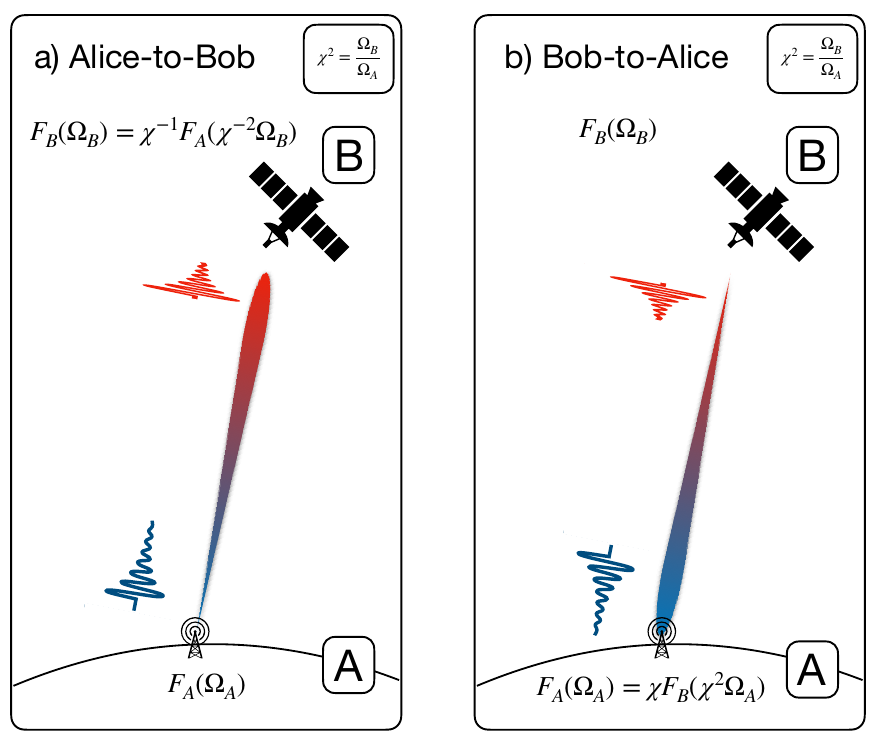}
    \caption{Depiction of the two possible scenarios: Alice-to-Bob and Bob-to-Alice photon exchange protocols. The effective transformations as proposed in the literature \cite{Bruschi:Ralph:2014} are presented for clarity. Note that the definition of the redshift $\chi^2=\Omega_\textrm{B}/\Omega_\textrm{A}$ is, and must be, the same in both cases.}
    \label{Figure:Clarification:1}
\end{figure}

It is now crucial to comment on the constitutive relation \eqref{fundamental:relation}. First, we note that this expression is correct as printed here, while it has been written incorrectly in parts of the literature on this topic \cite{Bruschi:Ralph:2014,Bruschi:Schell:2023,Alanis:Schell:Bruschi:2023}. In some of those works, the Alice-to-Bob transformation was reported as $F'(\omega)= \chi F(\chi^2\omega)$. Regardless, all quantities that were computed were unaffected by this problem: it is immediate to verify that quantity $|\langle F|F'\rangle|$ is independent of which way the transformation is written. We then note that the frequencies $\omega$ that appear in this relation are those measured locally by the receiver, Bob. This means that Bob is ideally in possession of two photon profiles $F(\omega)$ and $F'(\omega)$ for the purposes of the scheme at hand, which are defined locally in his frame. That is, the function $F$ has the same functional dependence on $\omega$ in both Alice's and Bob's laboratories, while $F'$ is the frequency profile that Bob receives from Alice. From now on, we assume that Alice is sending a photon to Bob and that the redshift is defined through \eqref{gravitational:redshift}. This implies that the mode transformation has the expression \eqref{fundamental:relation}. In case that we wish to discuss another process, this will be stated explicitly when necessary.  A depiction of the general setup is given in Figure~\ref{Figure:Clarification:1}.

\subsection{Gravitational redshift of quantum photons as a multimode mixer}
The set $\{F_{\vec{\zeta}}\}$ forms a complete orthonormal basis for the approximate one-dimensional model, as discussed above. The key aspect here is that the bases are infinite-dimensional. Since we are working in a free theory, we can choose two bases $\mathcal{B}=\{F_{\vec{\zeta}}\}$ and $\mathcal{B}'=\{F_{\vec{\zeta}'}'\}$ and write the Bogoliubov transformation between them in abstract form as
\begin{align}\label{Bogoliubov:transfromation:fat:modus}
    F_{\vec{\zeta}'}'=\int \textrm{d}\vec{\zeta}\left[\alpha_{\vec{\zeta}'\vec{\zeta}}F_{\vec{\zeta}}+\beta_{\vec{\zeta}'\vec{\zeta}}F_{\vec{\zeta}}^*\right],
\end{align}
where $\alpha_{\vec{\zeta}'\vec{\zeta}}:=\langle F_{\vec{\zeta}'}',F_{\vec{\zeta}}\rangle$ and $\beta_{\vec{\zeta}'\vec{\zeta}}:=-\langle F_{\vec{\zeta}'}',F_{\vec{\zeta}}^*\rangle$ are the Bogoliubov coefficients that satisfy the usual Bogoliubov identities \cite{Birrell:Davies:1982}.
The transformation between the two bases $\mathcal{B}$ and $\mathcal{B}'$ is \textit{linear}, i.e., each element of one basis can be written as a linear combination of elements of the other basis.\footnote{This linearity is unrelated to the inbuilt linearity of quantum mechanics.}
A paramount example of such transformation is the map between Minkowski and Unruh modes in flat Minkowski spacetime when studying the Unruh effect \cite{Unruh:1976,Bruschi:Louko:2010}. Linearity always arises from unitary transformations between operators that are generated by a quadratic Hamiltonian in the creation and annihilation operators themselves \cite{Zhang:Dong:2022}. This realization allows us to borrow from the powerful tools of symplectic geometry that have been extensively developed in the literature \cite{Adesso:Ragy:2014}. 

We assume for simplicity of presentation that the indices of the bases of interest are discrete, and therefore conveniently write the bases as $\mathcal{B}=\{F_n\}$ and $\mathcal{B}'=\{F_n'\}$. The operators $\hat{A}_n,\hat{A}_n^\dag$ and $\hat{A}_n',\hat{A}_n'{}^\dag$ of each basis  can be conveniently collected in the vectors $\hat{\mathbb{X}}:= ( \hat{A}_1, ...,\hat{A}_1^\dag,.... )^\textrm{Tp}$ and $\hat{\mathbb{X}}':= ( \hat{A}_1',...,\hat{A}_1'{}^\dag,...)^\textrm{Tp}$ respectively. Note that this discrete basis contains an infinite amount of elements. The commutation relations can be easily written as $[\hat{X}_n,\hat{X}_
m^\dag]=i\Omega_{nm}$ by introducing the \emph{symplectic form} $\boldsymbol{\Omega}:=\textrm{diag}(-i,...,i,...)$. This ultimately enables us to re-cast the  transformation \eqref{Bogoliubov:transfromation:fat:modus} as follows:
\begin{equation}\label{eq symplectic formalism}
    \hat{\mathbb{X}}'(\chi) := \hat{U}^\dagger (\chi)\, \hat{\mathbb{X}}\, \hat{U}(\chi)\equiv \boldsymbol{S}(\chi) \, \hat{\mathbb{X}}
    =
    \begin{pmatrix}
        \boldsymbol{\alpha}(\chi) & \boldsymbol{\beta}(\chi)\\
        \boldsymbol{\beta}^*(\chi) & \boldsymbol{\alpha}^*(\chi)
    \end{pmatrix}
    \hat{\mathbb{X}},
\end{equation}
where $\boldsymbol{S}(\chi)$ is the \emph{symplectic} matrix representation of the unitary operator $\hat{U}(\chi)$. A symplectic matrix $\boldsymbol{S}$ is defined as a matrix that satisfies the constraint $\boldsymbol{S}^\dag\boldsymbol{\Omega}\boldsymbol{S}=\boldsymbol{S}\boldsymbol{\Omega}\boldsymbol{S}^\dag=\boldsymbol{\Omega}$. Here $\boldsymbol{\alpha}(\chi)$ and $\boldsymbol{\beta}(\chi)$ are matrices that collect the respective Bogoliubov coefficients \cite{Adesso:Ragy:2014}.

We now note the following crucial aspect: in the regimes of interest, we expect a single photon to be sent and received under ideal conditions. This is in the nature of the redshift: the photon might be altered, but the number of photons sent is not affected \cite{Bruschi:Schell:2023,Alanis:Schell:Bruschi:2023}. Thus, it is natural to require that $\boldsymbol{\beta}(\chi)\equiv0$ since it is the $\beta$-coefficients that are responsible for the change of the number of particles, i.e., they are active transformations \cite{Adesso:Ragy:2014}. When this occurs, we note that the matrix $\boldsymbol{S}(\chi)$ is block diagonal and can be decomposed as $\boldsymbol{S}(\chi)=\boldsymbol{U}_\textrm{r}(\chi)\oplus\boldsymbol{U}_\textrm{r}^*(\chi)$, where $\boldsymbol{U}_\textrm{r}(\chi)$ is a unitary matrix acting on the reduced spaces $\hat{\mathbb{X}}_\textrm{r}:= ( \hat{A}_1, ...,)^\textrm{Tp}$ and $\hat{\mathbb{X}}_\textrm{r}^\dag:= ( \hat{A}_1^\dag, ...,)^\textrm{Tp}$ of annihilation or creation operators only. Thus, the independent part of \eqref{eq symplectic formalism} reduces to
\begin{equation}\label{eq symplectic formalism:reduced}
    \hat{\mathbb{X}}_\textrm{r}'(\chi) := \hat{U}^\dagger (\chi)\, \hat{\mathbb{X}}_\textrm{r}\, \hat{U}(\chi)\equiv \boldsymbol{U}_\textrm{r}(\chi) \, \hat{\mathbb{X}}_\textrm{r},
\end{equation}
where $\boldsymbol{U}(\chi)$ is the matrix representation of an appropriate unitary operator $\hat{U}(\chi)$. A second completely equivalent relation is obtained for $\hat{\mathbb{X}}_\textrm{r}^\dag$ from \eqref{eq symplectic formalism}, which we discard.

We now move on to the decomposition of the photonic space constructed above. A photon $|1_G\rangle\equiv\hat{A}_G^\dag|0\rangle$ with spectrum $G$ can be decomposed in spectrum-space on any basis. In particular, we can use $|1_{F_n}\rangle=\hat{A}_n^\dag|0\rangle$ and the decomposition of the identity in the one-particle sector $\mathds{1}=\sum_n |1_{F_n}\rangle\langle 1_{F_n}|$ to write
\begin{align*}
    |1_G\rangle=\sum_n \langle 1_{F_n}|1_G\rangle\, |1_{F_n}\rangle=&\sum_n C_n(G)|1_{F_n}\rangle.
\end{align*}
Since we choose single-photon states to be normalized by $\langle1_G|1_G\rangle=1$, we know that $\sum_n|C_n(G)|^2=1$. Let us now assume that we want to isolate the contribution of a particular mode function $F_n$ of interest in the expansion of $|1_G\rangle$. Without loss of generality we can select the contribution of the first vector with $n=1$. We thus write
\begin{align}
|1_G\rangle=&C_1(G)|1_{F_1}\rangle+\sum_{n\geq2} C_n(G)|1_{F_n}\rangle
=e^{i\phi}\cos\theta|1_{F_1}\rangle+
e^{i\psi}\sin\theta|1_\perp\rangle
\end{align}
since we have that $|C_1(G)|^2+\sum_{n\geq2}|C_n(G)|^2=1$,
and we have identified $C_1(G)\equiv e^{i\phi}\cos\theta$, the function $\sin^2\theta=\sum_{n\geq2}|C_n(G)|^2$, as well as the \textit{perpendicular mode}
\begin{equation*}
|1_\perp\rangle=|\sin\theta|^{-1}\sum_{n\geq2} C_n(G)|1_{F_n}\rangle.   
\end{equation*}
Such perpendicular mode is an artificial normalized (i.e., $\langle1_\perp|1_\perp\rangle=1$) mode that conveniently collects all elements of the basis that are not of interest to the problem. The phase $e^{i\phi}$ is in principle different from $1$ and is determined by the overlap of $|1_G\rangle$ and $|1_{F_1}\rangle$. It can be eliminated by absorbing it into the definition of $F_1$. We call this choice of decomposition the \emph{$(1+1)$-decomposition}.

The $(1+1)$-decomposition is the construction proposed, in essence, in the original work on this topic \cite{Bruschi:Ralph:2014}.
The next logical step is to extend it beyond the choice of one single mode of interest to an $(N+1)$-decomposition. In the literature \cite{Bruschi:Schell:2023,Alanis:Schell:Bruschi:2023}, it has been argued that the matrix $\boldsymbol{U}(\chi)$ can be constructed via the following simple procedure: one first selects $N$ modes of interest $\{F_1,...,F_N\}$ out of the basis to construct the operators $\hat{A}_n$, with $\langle F_j,F_k\rangle=\delta_{jk}$, and then extends the procedure just presented above to obtain
\begin{align}\label{redshift:matrix:transformation}
    \boldsymbol{U}(\chi)
    =
    \begin{pmatrix}
        U_{11}  & ... & U_{1N} & U_{1\perp}\\
        \vdots & \ddots & \vdots & ...\\
        U_{N1} & ... & U_{NN} & U_{N\perp}\\
        U_{\perp1} & ... & U_{\perp N} & U_{\perp\perp}
    \end{pmatrix},
\end{align}
where the following constitutive relations are required:
    \begin{equation}\label{constraints}
      U_{jk}:=\langle F_j,F_k'\rangle,
      \qquad
      \textrm{and}
      \qquad
      \sum_{p=1}^{N+1} U_{np}U_{mp}^*=\delta_{nm}.
      \end{equation}
It is easy to verify that the constraints above imply $|U_{n\perp}|^2=1-\sum_{p=1}^N |U_{np}|^2$, as well as $|U_{\perp\perp}|^2=1-\sum_{p=1}^N |U_{\perp p}|^2$.   
Note that, for the purpose of conveniently writing the sums above, we have used here the equivalent notation $F_\perp\equiv F_{N+1}$.

The transformation \eqref{redshift:matrix:transformation} is known in quantum optics as a \textit{multimode mixer}, or \textit{generalized beam-splitter} \cite{Adesso:Ragy:2014}. This transformation mixes linearly any amount of input modes, while conserving the total number of excitations. Its 
paramount incarnation is the celebrated \textit{beam-splitter}, which mixes two modes only \cite{Nielsen_Chuang:2010}. Note that the ``leak of information'' to the perpendicular mode $F_\perp$, which is assumed to be inaccessible for the purposes of the protocol, does not constitute true decoherence in the standard sense: in principle, if the total number of modes is low, the overall unitarity of the transformation guarantees that the information leaked away from the system can also flow back (at least in part). When the number of modes is high, a full rotation back into the system is less likely, although not impossible. In any case, as in all quantum optical processes, the degree and magnitude of oscillations between the subsystems greatly depend on the parameters of the system. For these reasons, we avoid claiming that this process involves, in its most theoretical incarnation, genuine decoherence.

We now make a comment regarding the applicability of linear dynamics. Linear dynamics are induced by Hermitian operators $\hat{H}$, e.g. Hamiltonians, that are quadratic in the creation and annihilation operators \cite{Adesso:Ragy:2014}. Concretely, we can write
\begin{align}
    \boldsymbol{S}=e^{\boldsymbol{\Omega}\boldsymbol{H}},
    \qquad
    \textrm{with}
    \qquad
    \boldsymbol{H}
    =
    \begin{pmatrix}
        \boldsymbol{U}  & \boldsymbol{V}\\
        \boldsymbol{V}^* & \boldsymbol{U}^*
    \end{pmatrix},
\end{align}
where $\boldsymbol{H}$ is the Hamiltonian matrix representation obtained from a Hamiltonian operator as $\hat{H}=\frac{1}{2}\hat{\mathbb{X}}^\dag\cdot\boldsymbol{H}\cdot\hat{\mathbb{X}}$, and $\boldsymbol{U}=\boldsymbol{U}^\dag$ and $\boldsymbol{V}=\boldsymbol{V}^\textrm{Tp}$.

The key transformation $\hat{\mathbb{X}}'=\boldsymbol{S}\hat{\mathbb{X}}$ transforms the modes contained in the vector $\hat{\mathbb{X}}$ as discussed before. This is particularly useful when the state $\hat{\rho}$ of the system is Gaussian (i.e., it has a Gaussian Wigner function \cite{Adesso:Ragy:2014}). In that case, the state is fully determined by its first and second moments that are collected in a vector $\vec{d}$ and covariance matrix $\boldsymbol{\sigma}$ respectively. When the state is transformed under linear dynamics one has $\vec{d}'=\boldsymbol{S}\vec{d}$ as well as $\boldsymbol{\sigma}'=\boldsymbol{S}\boldsymbol{\sigma}\boldsymbol{S}^\dag$. We leave a detailed introduction into the topic of Covariance Matrix formalism to the interested reader \cite{Adesso:Ragy:2014}.
The important observation that we wish to make is the following: while linear dynamics is particularly powerful in terms of the ability to obtain analytical expressions when combined with the covariance matrix formalism, linear dynamics applies in the way described above also for situations where the state of the system is not Gaussian. The main difference is that the evolution of the state cannot be obtained using a simple matrix relation $\boldsymbol{\sigma}'=\boldsymbol{S}\boldsymbol{\sigma}\boldsymbol{S}^\dag$, but one needs to compute the change in operators via $\hat{\mathbb{X}}'=\boldsymbol{S}\hat{\mathbb{X}}$ and use these expressions wherever needed.

\section{Characterizing gravitational redshift of quantum photons}\label{sec characterizing redshift}
In this section we provide an analysis of key aspects of QOGRM. These insights reveal properties that are necessary for the discussion of the main result of this work.

\subsection{Frequency change}
We start by asking the following question: \textit{what is the relation between the average frequency of the photon as measured locally by the receiver as compared to the average frequency of the photon as measured locally by the sender}? 
To answer our question we first define the \textit{average frequency} $\bar{\Omega}_K$ of the photon in mode $F$ as measured by observer K, with K$=$A,B for Alice and Bob respectively. Note that the mode $F$ is arbitrary, and is specified when necessary. Following the standard procedure of computing averages of a quantity $A(x)$ with probability distribution $p(x)$, we introduce $\bar{\Omega}_K$ via
\begin{align}
\bar{\Omega}_K=\int \textrm{d}\Omega_K\,\Omega_K\,|F(\Omega_K)|^2,
\end{align}
where $\Omega_K$ is the frequency as measured locally by Alice or Bob.

We then recall that the following are the constitutive relations of our problem:
\begin{align*}
    \frac{\Omega_\text{B}}{\Omega_\text{A}}=&\chi^2,
    &&\textbf{Redshift relation}\\
    F_\text{B}(\Omega_\text{B})=&\chi^{-1} F_\text{A}(\chi^{-2}\Omega_\text{B}),
    &&\textbf{Mode transformation}
\end{align*}
where the second expression represents the Alice-to-Bob scenario, while $F_\text{A}(\Omega_\text{A})=\chi F_\text{B}(\chi^2\Omega_\text{A})$ represents the opposite one (see Figure~\ref{Figure:Clarification:1}).

We first compute the average frequency $\bar{\Omega}_\text{A}$ of the mode $F_\text{A}(\Omega_\text{A})$, that is, the average frequency that Alice expects the agreed-upon mode to have. It reads 
    \begin{align*}
    \bar{\Omega}_\text{A}=&\int \textrm{d}\Omega_\text{A}\,\Omega_\text{A}\,|F_\text{A}(\Omega_\text{A})|^2.
    \end{align*}
We then continue by computing the average frequency $\bar{\Omega}^{\text{B}}_\text{A}$ of the mode $F_\text{A}(\Omega_\text{B})$, that is, the average frequency that Bob expects the agreed-upon mode to have. It reads 
    \begin{align*}
        \bar{\Omega}^{\text{B}}_\text{A}=&\int \textrm{d}\Omega_\text{B}\,\Omega_\text{B}\,|F_\text{A}(\Omega_\text{B})|^2\equiv\bar{\Omega}_\text{A}.
    \end{align*}    
    We conclude that the frequency $\bar{\Omega}^{\text{B}}_\text{A}$ computed by Bob in the mode $F_\text{B}(\Omega_\text{A})$
    coincides in value with the average frequency $\bar{\Omega}_\text{A}$ computed by Alice in the equivalent mode $F_\text{A}(\Omega_\text{A})$, that is, the average frequency that Alice assigns to the agreed-upon mode. 
    This occurs by construction, since Alice and Bob have agreed before the protocol on a particular mode, namely $F_{\text{A}}$, that they use as reference. From now on we use $\bar{\Omega}_\text{A}$ as the reference average frequency.
    
    Bob then proceeds to compute the average frequency $\bar{\Omega}_\text{B}$ of the received photon. This reads
\begin{align*}
    \bar{\Omega}_\text{B}=&\int \textrm{d}\Omega_\text{B}\,\Omega_\text{B}\,|F_\text{B}(\Omega_\text{B})|^2=\chi^{-2}\int \textrm{d}\Omega_\text{B}\,\Omega_\text{B}\,|F_\text{A}(\chi^{-2}\Omega_\text{B})|^2
    =\chi^2\int \textrm{d}\Omega_\text{B}'\,\Omega_\text{B}'\,|F_\text{A}(\Omega_\text{B}')|^2=\chi^2\bar{\Omega}_\text{A}.
\end{align*}
Thus, we have found
\begin{align}\label{average:frequency:gravitational:redshift}
\bar{\Omega}_\text{B}=&\chi^2\bar{\Omega}_\text{A},
\end{align}
which agrees with the first constitutive relation and is the expected transformation for the gravitational redshift of photonic frequencies. Notice that if we were to consider the Bob-to-Alice scenario and compute the average frequency $\bar{\Omega}_\text{A}$, we would have obtained $\bar{\Omega}_\text{A}=\chi^{-2}\bar{\Omega}_\text{B}$, which is equivalent to \eqref{average:frequency:gravitational:redshift} as expected.

\begin{figure}[t!]
    \centering
    \includegraphics[width=0.8\linewidth]{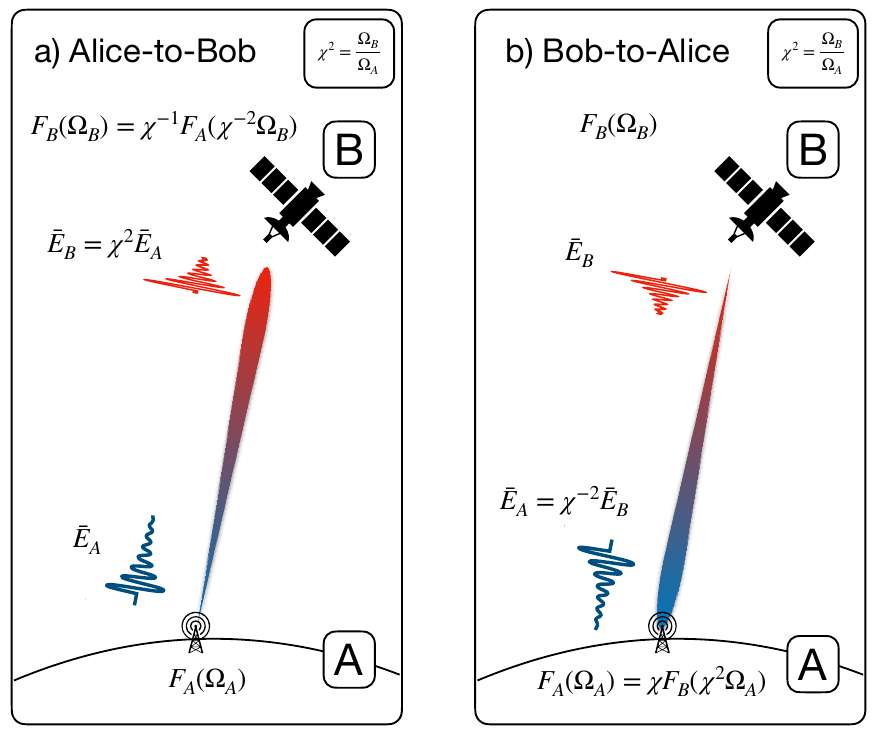}
    \caption{Depiction of the two possible scenarios: Alice-to-Bob and Bob-to-Alice photon exchange protocols. The effective transformations as proposed in the literature \cite{Bruschi:Ralph:2014} are presented for clarity, together with the changes in average energy.}
    \label{Figure_Clarification_2_Caption}
\end{figure}

We now note that the energy $E_K$ of a photon with frequency $\Omega_K$ as measured locally by Alice or Bob is given by $E_K=\hbar\Omega_K$, where we have restored Planck's constant here for the sake of the argument. We can use this expression to compute the change in average energy $\bar{E}_K$ as measured locally by Bob in the Alice-to-Bob scenario. We have
\begin{align}
\bar{E}_\text{B}=&\chi^2\bar{E}_\text{A},
\end{align}
again, as expected.
We can therefore conclude this brief analysis by computing the variation in energy $\Delta\bar{E}:=\bar{E}_\text{B}-\bar{E}_\text{A}$ \textit{as measured locally} by Bob. It reads $\Delta\bar{E}=(\chi^2-1)\bar{E}_\text{A}=z \bar{E}_\text{A}$, where we have restored the standard notation $z$ for the redshift for completeness. Figure~\ref{Figure_Clarification_2_Caption} gives a pictorial summary of this result.

\subsection{Perturbative expansion of the overlap}
We now proceed to study the perturbative behaviour of the overlap 
\begin{align}\label{delta:definition}
    \Delta(\chi):=\langle F| F'\rangle=\chi^{-1}\int \mathrm{d}x F^*(x)F(\chi^{-2}x)=\chi\int \mathrm{d}x F^*(\chi^2x)F(x)
\end{align}
of the mode functions $F,F'$ in the regimes $|\chi-1|\ll1$ and $\chi\gg1$, which will aid our understanding of the problem at hand.\footnote{Note that the regime $\chi\ll1$ can be obtained immediately from the second case under consideration by a change of variables.}
This quantity is defined in the Alice-to-Bob scenario, and is determined by the constraints: $\int \mathrm{d}\omega |F(\omega)|^2=\int \mathrm{d}\omega |F'(\omega)|^2=1$ and  $F'(\omega)=\chi^{-1} F(\chi^{-2}\omega)$. Note that $|\Delta|\leq \sqrt{\langle F|F\rangle}\sqrt{\langle F'|F'\rangle}=1$ for all $F,F'$. The function $F(\omega)$ will include at least one constant of dimension frequency that dictates the overall effective width of the function, i.e., the function must be mostly contained within a finite interval since it is a function in $L^2$, which decreases as $\omega$ increases. We are making the assumption that the function is effectively defined only on the positive $\omega$ axis, as is standard in quantum field theory \cite{Srednicki:2007}. In the following we will use the notation $x:=\omega/\sigma$, where $x$ will be a dimensionless variable and $\sigma$ determines the overall bandwidth of the mode. Therefore, we will replace the dimensionful functions $F(\omega)$ (which have dimension since $\int \mathrm{d}\omega |F(\omega)|^2=1$) with the dimensionless functions $F(x)=\sqrt{\sigma}F(x=\omega/\sigma)$, as already done in \eqref{delta:definition}.

\subsubsection{Perturbative expansion of the overlap: $|\chi-1|\ll1$}
Here we consider the case of small effects where $\chi-1=:\epsilon\ll1$ and $\epsilon>0$ implies blueshift. The case of redshift applies in a straightforward fashion. We wish to compute $\Delta(\chi)$ defined in \eqref{delta:definition} using a Taylor series around the no-redshift case $\chi=1$. We have
\begin{align}\label{mode:function:Taylor:series}
    F(\chi^2x)\simeq F(x)+2 x \dot{F}(x) \epsilon+\left(x \dot{F}(x)+2 x^2 \ddot{F}(x)\right)\epsilon^2, 
\end{align}
and we must assume that the functions are well behaved such that the coefficients at each order do not grow without bound. 
This expression can be inserted in \eqref{delta:definition} with $\Delta_\epsilon:=\Delta(\chi=1+\epsilon)$, and we obtain
\begin{align} \label{delta epsilon}
    \Delta_\epsilon=1+\epsilon+2 \mathcal{K}\epsilon- (3\mathcal{K}+2\mu^2)\epsilon^2,
\end{align}
where we have defined the constitutive quantities
\begin{subequations}
    \begin{align}
        \mathcal{K}:=&\int \mathrm{d}x\,x F^*(x) \dot{F}(x),\\
        \mu^2:=&\int \mathrm{d}x\,x^2 \dot{F}(x)\dot{F}^*(x),
    \end{align}
\end{subequations}
for convenience of presentation. Note that $\mu^2$ is purely real. Furthermore we can deduce that $\mathcal{K}+\mathcal{K}^*=2\Re(\mathcal{K})=-1$, since $\mathcal{K} =\int \mathrm{d}x\, x F^*(x) \dot{F}(x) = [x F^*(x) F(x)]_0^{\infty} - \int \mathrm{d}x \, F^*(x) F(x) - \int \mathrm{d}x \, x \dot{F}^*(x) F(x) = -1-\mathcal{K}^* $. Here, the fact that the boundary term vanishes at infinity is guaranteed by the fast decrease of the function $F$. Therefore, we have shown that
\begin{equation*}
    \Re(\mathcal{K})=-\frac{1}{2}.
\end{equation*}
We write $\mathcal{K}= -\frac{1}{2}+i\kappa$, where $\kappa\equiv\Im(\mathcal{K})$ is real, which allows us to simplify \eqref{delta epsilon} to
\begin{equation}\label{delta:perturbative:expansion}
    \Delta_\epsilon= 1+2i\epsilon\kappa+\frac{\epsilon^2}{2}-3i\epsilon^2\kappa-2\epsilon^2\mu^2.
\end{equation}
Notice that a vanishing first order real contribution to $\Delta_\epsilon$ is consistent with $|\Delta_\epsilon|\leq1$ for all $\epsilon$.

We then proceed to compute the real and imaginary parts $\Re(\Delta_\epsilon)$ and $\Im(\Delta_\epsilon)$ respectively of the overlap $\Delta_\epsilon$. We find
\begin{align*}
\Re(\Delta_\epsilon)=1+\frac{\epsilon^2}{2}\left(1-4\mu^2\right),\quad\quad\Im(\Delta_\epsilon)=2\epsilon\kappa-3\epsilon^2\kappa.
\end{align*}
We are finally able to compute the perturbative expression of the magnitude of $\Delta_\epsilon$, and to lowest significant order find
\begin{equation}
    |\Delta_\epsilon|=1-\frac{\epsilon^2}{2}\left(4\mu^2-4\kappa^2-1\right).
\end{equation}
Notice that $|\Delta_\epsilon|\leq1$ by construction, which implies that $\mu^2\geq\kappa^2+\frac{1}{4}$.

We can put together these results to conveniently write
\begin{equation}\label{delta:exponential:form}
\Delta_\epsilon=e^{2i\kappa(1-\frac{\epsilon}{2})\epsilon}e^{-\frac{\epsilon^2}{2}\left(4\mu^2-4\kappa^2-1\right)},
\end{equation}
which holds to second order in $\epsilon$. 

We can derive another expression for the function $\kappa$, which helps us to better appreciate its importance. We start by recalling that $\kappa\equiv\Im(\mathcal{K})$. We then write $F(x)=e^{i\theta(x)}\sqrt{\rho(x)}$, where $\rho(x)=|F(x)|^2$. We can repeat the simple computations done above and obtain:
\begin{subequations}\label{alternative:kappa:expression}
\begin{alignat}{4}
\kappa=&\int \mathrm{d}x\,x\,\dot{\theta}(x) \rho(x),\\
\mu^2=&\int \mathrm{d}x\,x^2\,\left(\dot{\theta}^2(x)+\frac{\dot{\rho}^2(x)}{\rho^2(x)}\right) \rho(x).
\end{alignat}
\end{subequations}
The expressions \eqref{delta:exponential:form} and \eqref{alternative:kappa:expression} immediately give us insight into the role that the phase $\theta$ of the wavepacket and its magnitude $\sqrt{\rho}$ play in the expression \eqref{delta:exponential:form}: the phase shift $e^{2i\kappa(1-\frac{\epsilon}{2})\epsilon}$ is crucially dependent on the change in the phase $\theta$, while the magnitude $|\Delta_\epsilon|=e^{-\frac{\epsilon^2}{2}\left(4\mu^2-4\kappa^2-1\right)}$ of the wavepacket (and therefore of its deformation) is determined by the change of both $\theta$ and $\rho$. The expressions \eqref{alternative:kappa:expression} can also be combined to show that $\mu^2-\kappa^2=(\Delta\sigma_{x\dot{\theta}})^2+\int \mathrm{d}x\,x^2\,\frac{\dot{\rho}^2(x)}{\rho(x)}\geq0$, where $(\Delta\sigma_{g(x)})^2:=\int \mathrm{d}x\,g^2(x) \rho(x)-(\int \mathrm{d}x\,g(x) \rho(x))^2\geq0$ is the variance of the quantity $g(x)$ in the normal distribution $\rho(x)$.

We note in passing that the fact that the magnitude of the deformation changes at second order is consistent with the preliminary results already found in the literature \cite{Bruschi:Chatzinotas:2021}.
This implies that, to first order, the effects of a small gravitational redshift on the wavepacket of a photon effectively correspond to the presence of an overall additional phase factor to the wavepacket.

\subsubsection{Asymptotic analysis of the overlap: $\chi\gg1$ or $\chi\ll1$}
Here we are interested to show that $|\Delta(\chi)|\rightarrow0$ for both limits $\chi\rightarrow\infty$ and \(\chi\rightarrow 0\). To tackle this goal we introduce the overlaps $\Delta_{nm}(\chi):=\langle F_n|F_m'\rangle$ between two mode functions $F_n$ and $F_m$, which have the explicit expression
\begin{align}
    \Delta_{nm}(\chi)=&\chi^{-1}\int_{0}^{+\infty} \mathrm{d}\omega F_n^*(\omega)F_m(\chi^{-2}\omega)=\chi\int_{0}^{+\infty} \mathrm{d}\omega F_n^*(\chi^2\omega)F_m(\omega).
\end{align}
We then recall that $|\int \mathrm{d}x f(x) |\leq\int \mathrm{d}x |f(x)|$ and we assume that $F_k$ are $L^1$ continuous functions that must therefore have a maximum $F_{\textrm{max},k}$ in the domain.

We now provide the following two bounds:
\begin{subequations}\label{overlap:bounds}
\begin{align}
|\Delta_{nm}(\chi)|=&\chi\left|\int_{0}^{+\infty} \mathrm{d}\omega F_n^*(\omega)F_m(\chi^2\omega)\right|\leq \chi\,F_{\textrm{max},m}\int_0^{+\infty} \mathrm{d}\omega |F_n(\omega)|,\label{overlap:bound:small}\\
|\Delta_{nm}(\chi)|=&\chi^{-1}\left|\int_{0}^{+\infty} \mathrm{d}\omega F_n^*(\omega)F_m(\chi^{-2}\omega)\right|\leq \chi^{-1}\,F_{\textrm{max},m}\int_0^{+\infty} \mathrm{d}\omega |F_n(\omega)|\label{overlap:bound:large},
\end{align}
\end{subequations}
which vanish in the limit of \(\chi \rightarrow 0\) and \(\chi \rightarrow +\infty\) respectively.
These simple expressions allow us to conclude that the overlap between a mode function and its transformed due to redshift is expected to vanish when the redshift is asymptotically large, or asymptotically small.

\section{On the domain of validity of gravitational redshift as a multimode mixer}\label{sec new work}
The QOGRM model reviewed above does not assume nor predict any constraint on the redshift itself. Therefore, the matrix representation $\boldsymbol{U}(\chi)$ given in \eqref{redshift:matrix:transformation} is assumed to be valid in the whole domain $0<\chi$.

In the following we present two simple yet illustrative examples that, together, challenge this implicit assumption. 
The idea is to construct the matrices $\boldsymbol{U}(\chi)$ for a different number of modes and verify if and when does this construction fail.

\subsection{Matrix representation of the gravitational redshift channel: $1+1$-decomposition}
We start by considering one mode $F$ of interest. The matrix $\boldsymbol{U}(\chi)$ that represents the transformation $F\rightarrow F'$ induced by the gravitational redshift reads
\begin{align}\label{redshift:one:mode:matrix:transformation}
    \boldsymbol{U}(\chi)
    =
    e^{i\phi}\begin{pmatrix}
        \cos\theta  & e^{i\varphi_1}\sin\theta\\
        -e^{i\varphi_2}\sin\theta & e^{i(\varphi_1-\varphi_2)}\cos\theta
    \end{pmatrix},
\end{align}
where $\theta=\theta(\chi)$, $\varphi_j=\varphi_j(\chi)$, and we have defined the angle $\theta$ via $\cos\theta:=|\langle F,F'\rangle|$ as prescribed above. Note that the phase $e^{i\phi}$ of $\Delta(\chi)=\langle F,F'\rangle$ has been collected as a global phase.
The fact that $\cos\theta\geq0$ in this case restricts the domain of the angle to $0\leq\theta(\chi)\leq\pi/2$. Notice also that the map $\chi\rightarrow\theta(\chi)$ is continuous but does not need to be injective. This is expected to happen if the mode $F$ has more than one local maximum, which occurs, for example, when considering frequency-comb spectral profiles \cite{nisbet:2011,rambach:2016}.

The matrix \eqref{redshift:one:mode:matrix:transformation} is always well defined and does not pose any challenge to the validity of the model. In particular, we can compute it explicitly in the two limits $\chi-1=\epsilon \ll1$ and $\chi\gg1$. It is easy to use the computations given above to obtain
\begin{equation}\label{very:useful:matrix}
    \boldsymbol{U}(1+\epsilon) \approx \mathds{1}+\begin{pmatrix}
 i u^{(1)}_{11} & u^{(1)}_{1\perp} \\
 -u^{(1)}_{1\perp}{}^* & i u^{(1)}_{\perp\perp}
\end{pmatrix}\epsilon,
\qquad
\text{and}
\qquad
\boldsymbol{U}(\chi\gg1) \approx \begin{pmatrix}
 0 & e^{i\varphi_{1\perp}} \\
 e^{i\varphi_{\perp1}} & 0
\end{pmatrix},
\end{equation}
where $u^{(1)}_{11}$,$u^{(1)}_{1\perp}$,and $u_{\perp\perp}$ are real numbers, and the explicit values of the quantities $u$, as well as of the phases $\varphi$, are irrelevant for our purposes here.

\subsection{Matrix representation of the gravitational redshift channel: $2+1$-decomposition}
We now move on to the two-mode case. We select modes $F_1$ and $F_2$ of interest, and once more construct the corresponding matrix $\boldsymbol{U}(\chi)$. We have
\begin{align}\label{redshift:two:mode:matrix:transformation}
    \boldsymbol{U}(\chi)
    =
    \begin{pmatrix}
        U_{11}  & U_{12} & U_{1\perp}\\
        U_{21}  & U_{22} & U_{2\perp}\\
        U_{\perp1}  & U_{\perp2} & U_{\perp\perp}
    \end{pmatrix},
\end{align}
where the elements of this matrix satisfy the constraints \eqref{constraints}.

\begin{figure}[t!]
    \centering
    \includegraphics[width=0.8\linewidth]{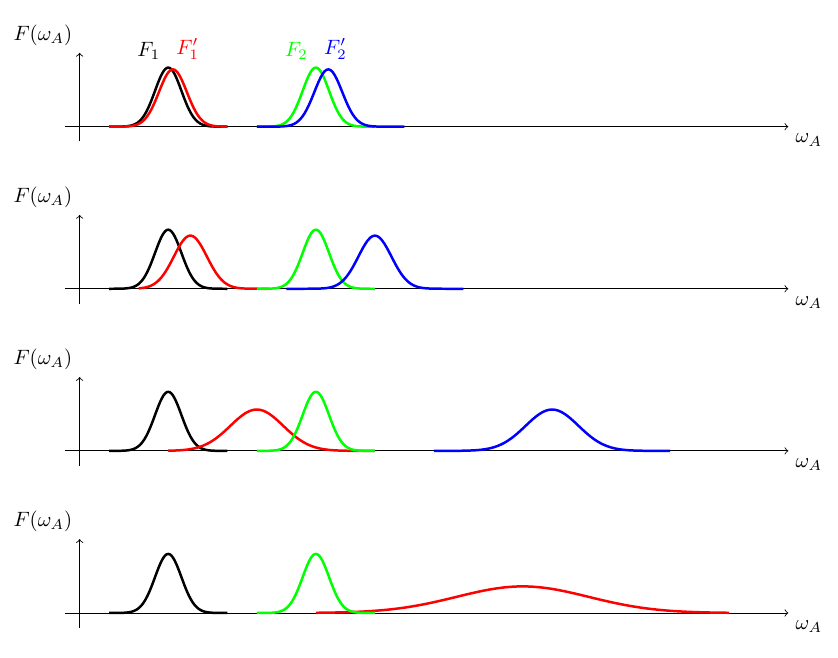}
    \caption{\textbf{Bob-to-Alice scenario}: Sketch of the transformation of two modes \(F_1\) and \(F_2\) into the modes \(F_1'\) and \(F_2'\), respectively. We have chosen the Bob-to-Alice scenario with $\chi \leq 1$ since the modes move to the right in the frequency domain as $\chi$ increases, which allows for better visualisation. Note that \(F_2'\) has moved outside of the figure in the last panel. 
    It is clear from the figure that, with increasing redshift, the overlap of all transformed modes $F_n'$ with any of the initial modes $F_n$ diminishes, and eventually vanishes. All frequencies are to be measured in the \textit{local} reference frame of Alice.}
    \label{fig:mode:transformation}
\end{figure}

The key observation is that the elements $U_{nm}\equiv\langle F_n|F_m'\rangle$. In particular, we have $F_n'(\omega)=\chi^{-1}\,F_n(\chi^{-2}\,\omega)$ in the Alice-to-Bob scenario, and we have already shown above what occurs to the overlap of two functions related by this transformations when $|\chi-1|\ll1$ and $\chi\rightarrow+\infty$. We use the results of the previous sections, and the constraints \eqref{constraints}, to obtain the asymptotic expressions for the matrix \eqref{redshift:two:mode:matrix:transformation}, which read
\begin{align}\label{redshift:two:mode:matrix:transformation:extreme}
    \boldsymbol{U}(1+\epsilon)
    \approx
    \mathds{1}+\begin{pmatrix}
        i u_{11}  & u_{12} & u_{1\perp}\\
        -u^*_{12}  & i u_{22} & u_{2\perp}\\
        -u_{1\perp}^*  & -u_{2\perp}^* & i u_{\perp\perp}
    \end{pmatrix}\epsilon,
    \quad
    \boldsymbol{U}(\chi\gg1)
    \approx
    \begin{pmatrix}
        0  & 0 & e^{i\varphi_{1\perp}}\\
        0  & 0 & e^{i\varphi_{2\perp}}\\
        e^{i\varphi_{\perp1}}  & e^{i\varphi_{\perp2}} & U_{\perp\perp}
    \end{pmatrix}.
\end{align}
We here stumble upon the problem whose presence has been foreshadowed above. The matrix \eqref{redshift:two:mode:matrix:transformation:extreme} cannot be unitary for all of the values of $\chi$, as is obvious from its form for $\chi\gg1$. Figure~\ref{fig:mode:transformation} provides a visual understanding of the $2+1$-decomposition.

\subsection{Considerations on the domain of validity}
We have presented the two simplest possible applications of QOGRM, namely the $1+1$ and $2+1$-decompositions. We have seen that the $1+1$ case does not pose any problem, while already in the $2+1$ case, the decomposition does not hold for all values $\chi^2$ of the redshift. 

In order to understand why this occurs it is necessary to take a step back and consider how this model was constructed in the first place. Let us start by looking at the perturbative regime around $\chi=1$. Since we are trying to construct a unitary matrix $\boldsymbol{U}(\chi)$, in the regime $\chi=1+\epsilon$ where $\boldsymbol{U}(\chi=1)=\mathds{1}$ it is immediate to see that $\boldsymbol{U}(\chi)$ has perturbative expansion
\begin{align}
    \boldsymbol{U}(\chi=1+\epsilon)=\mathds{1}+\boldsymbol{U}^{(1)}\epsilon,
\end{align}
where $\boldsymbol{U}^{(1)}{}^\dag=-\boldsymbol{U}^{(1)}$. This implies that the elements $u_{nm}$ of $\boldsymbol{U}^{(1)}$ must satisfy the following: $u_{nn}$ are purely imaginary, while $u_{mn}=-u_{nm}^*$ for $n\neq m$. 

We then note that, according to our construction, we must have the identifications $U_{nm}=\Delta_{nm}\equiv\langle F_n|F'_m\rangle$. Since we are working in the small redshift perturbative regime, we have $\Delta_{nm}=\delta_{nm}+\Delta^{(1)}_{nm}$, and therefore our unitary-matrix constraint becomes
\begin{align}\label{first:order:constraints}
    u_{mn}=-u_{nm}^*
    \qquad
    \Leftrightarrow
    \qquad
    \Delta^{(1)}_{mn}=-\Delta^{(1)}_{nm}{}^*.
\end{align}
We now write the explicit expressions for $\Delta_{nm}$ and $\Delta_{mn}$ below:
\begin{align*}
    \Delta_{nm}(\chi)=&\chi^{-1}\int_0^{+\infty}\textrm{d}\omega F_n^*(\omega)F_m(\chi^{-2}\omega),\\
    \Delta_{mn}(\chi)=&\chi^{-1}\int_0^{+\infty}\textrm{d}\omega F_m^*(\omega)F_n(\chi^{-2}\omega)=\chi\int_0^{+\infty}\textrm{d}\omega F_m^*(\chi^2\omega)F_n(\omega).
\end{align*}
The crucial observation at this point is that $\Delta_{nm}(\chi=1)=0$ for all $n\neq m$ by construction. Therefore, we expect that $\Delta_{nm}(\chi=1+\epsilon)=\Delta^{(1)}_{nn}\epsilon$ to lowest nontrivial order. We therefore employ $\chi=1+\epsilon$ in the two expressions above. The computations are given for the first expression, while the second is derived analogously. We have 
\begin{align*}
    \Delta_{nm}(\chi=1+\epsilon)
    =&(1+\epsilon)^{-1}\int_0^{+\infty}\textrm{d}\omega F_n^*(\omega)F_m((1+\epsilon)^{-2}\omega)\\
    =&(1+\epsilon)\int_0^{+\infty}\textrm{d}\omega F_n^*((1+\epsilon)^2\omega)F_m(\omega)\\
    =&(1+\epsilon)\int_0^{+\infty}\textrm{d}\omega F_n^*((1+2\epsilon)\omega)F_m(\omega)\\
    =&\epsilon\int_0^{+\infty}\textrm{d}\omega F_n^*(\omega)F_m(\omega)+2\epsilon\int_0^{+\infty}\textrm{d}\omega \,\omega\dot{F}_n^*(\omega)F_m(\omega)\\
    =&2\epsilon\int_0^{+\infty}\textrm{d}\omega \,\omega\dot{F}_n^*(\omega)F_m(\omega),\\
    \Delta_{mn}(\chi=1+\epsilon)=&-2\epsilon\int_0^{+\infty}\textrm{d}\omega \,\omega\dot{F}_n(\omega)F_m^*(\omega),
\end{align*}
where to obtain the fourth line we have employed the perturbative expansion of the mode functions around $\chi=1$ given in \eqref{mode:function:Taylor:series}, while to obtain the fifth line we have noted that the first term in the fourth line vanishes identically since it is equal to $\epsilon\langle F_n|F_m\rangle\equiv0$.
The computations above confirm that we can always have the constraints \eqref{first:order:constraints} satisfied for all $n,m$, including the diagonal terms. Therefore, QOGRM can always be applied to first order in the small redshift.

\vspace{0.2cm}

The problem with increasing redshift to a finite, perhaps large, amount is that the mode-overlaps must become very small in the asymptotic limit of infinitely large red- and blueshift as shown in \eqref{overlap:bounds}. 
To see how this leads to a problem, consider that the $N+1$-decomposition of interest is implemented by the matrix $\boldsymbol{U}(\chi)$ defined in \eqref{redshift:matrix:transformation} with constraints \eqref{constraints} that require that $\sum_{p=1}^N|U_{np}|^2=1-|U_{n\perp}|^2$ as well as $\sum_{p=1}^N|U_{p\perp}|^2+|U_{\perp\perp}|^2=1$. These two constraints clearly fail when $\chi$ is large enough. In fact, when $\chi\gg1$ we have found in \eqref{overlap:bound:large} that $|\Delta_{nm}|\leq\chi^{-1}C_{nm}$ for an appropriate constant $C_{nm}$ independent of $\chi$. Thus, in this regime one has 
\begin{align*}
    \sum_{n=1}^N|U_{n\perp}|^2=&N-\sum_{n=1}^N|U_{n\perp}|^2\geq2-\sum_{n,p=1}^N\chi^{-2}|C_{np}|^2\geq N-\chi^{-2}C,
\end{align*}
which can be made arbitrarily close to $N$, and therefore strictly larger than $1$ for $N\geq2$, since $C:=\sum_{n,p=1}^N|C_{np}|^2$ is independent of $\chi$. 

This, in turn, implies that $1=\sum_{p=1}^N|U_{p\perp}|^2+|U_{\perp\perp}|^2\geq N+|U_{\perp\perp}|^2-\chi^{-2}C>1$ for appropriate choices of $\chi$, which is impossible, and therefore QOGRM ceases to be applicable. The last strict inequality comes from the fact that one has always a value $\tilde{\chi}\gg1$ such that $|U_{\perp\perp}|^2-\chi^{-2}C>0$ for all $\chi>\tilde{\chi}$. An example where this issue is manifest is presented in the box below.

\begin{tcolorbox}[breakable, colback=teal!2!white,colframe=teal!85!white,title=Example: vanishing of the overlap of mode-functions with finite bandwidth]
 
One situation where this becomes a concrete issue that can be controlled analytically is that of functions with finite bandwith.
 In fact, let $F_k(\omega)=\textrm{Rect}((\omega-\omega_{0,k})/\sigma_k)f(\omega)$, where $f(\omega)$ is an $L^1$ function. Then, one has
 {\small
 \begin{align*}
    |\Delta_{nm}(\chi)|=&\chi^{-1}\left|\int_{0}^{+\infty} \mathrm{d}\omega F_n^*(\omega)F_m(\chi^{-2}\omega)\right|\\
    =&\chi^{-1}\left|\int_{0}^{+\infty} \mathrm{d}\omega\,
    \textrm{Rect}\left(\frac{\omega-\omega_{0,n}}{\sigma_n}\right)\textrm{Rect}\left(\frac{\chi^{-2}\omega-\omega_{0,m}}{\sigma_m}\right)
     f^*(\omega)f(\chi^{-2}\omega)\right|\\
     =&\chi^{-1}\left|\int_{0}^{+\infty} \mathrm{d}\omega\,
    \textrm{Rect}\left(\frac{\omega-\omega_{0,n}}{\sigma_n}\right)\textrm{Rect}\left(\frac{\omega-\chi^2\omega_{0,m}}{\chi^2\sigma_m}\right)
     f^*(\omega)f(\chi^{-2}\omega)\right|,
\end{align*}
}
which does not vanish as long as either $\omega_{0,n}+\sigma_n/2>\chi^2\omega_{0,m}-\chi^2\omega_m/2$ or $\chi^2\omega_{0,m}+\chi^2\omega_m/2>\omega_{0,n}-\sigma_n/2$. In other words, as long as there is a finite overlap between the intervals in which the rectangle functions are non vanishing one has $|\Delta_{nm}(\chi)|\neq0$.
Unfortunately, it is immediate to see that, for each condition, there is always a threshold value of $\chi_{nm}$ above which the constraints are violated. In such case, the overlaps $|\Delta_{nm}(\chi)|$ vanish identically, and it is sufficient to choose $\chi>\max\{\chi_{nm}\}$ to obtain vanishing overlaps for all modes of interest. This gives us a matrix $\boldsymbol{U}(\chi)$ of the form found in \eqref{redshift:two:mode:matrix:transformation:extreme} for large redshifts, which violates the prerequisite of the model itself.
\end{tcolorbox}

\section{Extension of the model to the full redshift domain}\label{section:solution}
We have shown that QOGRM is a model that can always be applied for small redshifts, but ceases to be applicable within $N+1$-decompositions with $N\geq2$ when the redshift is large enough. The natural question to ask at this point is if there is a fundamental breakdown of the physics of interest, or if the model can be amended to include all values of the redshift. 

Before concluding that the physics of interest is indeed not captured by the common understanding of the process at hand, we assume that QOGRM must be amended and put forward the following intuition behind the breakdown of the model: the $N+1$-decomposition initially proposed in QOGRM requires an infinite amount of ``environment'' modes to be collected in a ``perpendicular'' one. However, given the nature of the transformation \eqref{fundamental:relation}, and in particular its ``rigidity'' in terms of the physically required constraints, we conclude that the transformation effectively maps all chosen modes to the single existing perpendicular mode for large enough values of the redshift, which breaks down unitarity (except when $N=1$).

\subsection{Solution to the problem: general approach}

The most straightforward solution to the problem is to extend the matrix $\boldsymbol{U}(\chi)$ to include all of the basis $\mathcal{B}=\{F_n\}$, thereby taking the formal expression
\begin{align}
    \boldsymbol{U}(\chi)
    =
    \begin{pmatrix}
        U_{11}  & ... & U_{1N} & ...\\
        \vdots & \ddots & \vdots & \vdots\\
        U_{N1} & ... & U_{NN} & ...\\
        \vdots & \vdots & \vdots & \ddots
    \end{pmatrix},
\end{align}
which implies that it is an infinite-dimensional matrix where $U_{nm}\equiv\Delta_{nm}=\langle F_n|F_m'\rangle$ as mentioned before.

This immediately poses the question of rigorously defining such an object as well as how to determine all of its entries in practical terms. In fact, one can in principle construct a basis $\mathcal{B}=\{F_n\}$ and then compute the transformed basis $\mathcal{B}'=\{F_n'\}$, as well as all of the overlaps $\Delta_{nm}$. It is the clear that, regardless of the technical issue of obtaining all expressions of interest, QOGRM can in this case be implemented by a formally unitary matrix for all values of the redshift. This solution is not satisfactory since it requires us to concretely deal with infinite-dimensional spaces of matrices, and therefore we proceed below to provide an improvement on the proposed solution.

\subsection{Partial solution to the problem: extended decomposition}
The solution proposed above formally solves the problem identified by this work. This is, however, an unsatisfactory solution for concrete purposes, since it requires the computation of an infinite amount of quantities as already mentioned.
We can borrow from the intuition provided above and from the fact that the $1+1$-decomposition does not pose any challenges, to seek for an ``intermediate'' solution. We therefore extend the $N+1$-decomposition to and $N+M$-decomposition, with the idea of leaking the information of the $N$ modes of interest to additional $M$ effective perpendicular modes in such a way as to maintain the whole transformation unitary. This approach, if successful, would have the added benefit of providing a well-defined unitary transformation $\boldsymbol{U}(\chi)$, since the matrix would be finite-dimensional, at the arguably acceptable cost of requiring $M$ perpendicular modes whose exact properties one must then obtain through alternative means. 

To obtain our goal let us introduce the $(N+M)\times(N+M)$ matrix
\begin{align}\label{solution:matrix}
    \boldsymbol{U}(\chi)
    =
    \begin{pmatrix}
        U_{11}  & ... & U_{1N} & U_{1\perp_1} & ... & U_{1\perp_M}\\
        \vdots & \ddots & \vdots & \vdots & \ddots & \vdots\\
        U_{N1} & ... & U_{NN} &  U_{N\perp_1} & ... & U_{N\perp_M}\\
        U_{\perp_11} & ... & U_{\perp_1N} &  U_{\perp_1\perp_1} & ... & U_{\perp_N\perp_M}\\
        \vdots & \ddots & \vdots & \vdots & \ddots & \vdots \\
        U_{\perp_M1}  & ... & U_{\perp_M N} & U_{\perp_M\perp_1} & ... & U_{\perp_M\perp_M}
    \end{pmatrix},
\end{align}
where the entries of the matrix read $U_{nm}=\langle F_n|F_m'\rangle$, $U_{\perp_nm}=\langle F_{\perp_n}|F_m'\rangle$, and $U_{\perp_n\perp_m}=\langle F_{\perp_n}|F_{\perp_m}'\rangle$, and they are defined by the following constraints
\begin{subequations}\label{solution:rows:columns}
\begin{align}
    R^\textrm{u}_n:=&\sum_{p=1}^N|U_{np}|^2+\sum_{p=1}^M|U_{n\perp_p}|^2=1,\\
    R^\textrm{d}_n:=&\sum_{p=1}^N|U_{\perp_np}|^2+\sum_{p=1}^M|U_{\perp_n\perp_p}|^2=1,\\
    C^\textrm{l}_p:=&\sum_{p=1}^N|U_{pn}|^2+\sum_{p=1}^M|U_{\perp_p n}|^2=1,\\
    C^\textrm{r}_p:=&\sum_{p=1}^N|U_{p\perp_n}|^2+\sum_{p=1}^M|U_{\perp_p\perp_n}|^2=1,
\end{align}
\end{subequations}
in order for the matrix $\boldsymbol{U}(\chi)$ to be unitary.

We then compute the two following combinations of constraints $\sum_{n=1}^N R^\textrm{u}_n$ and $\sum_{n=1}^M R^\textrm{d}_n$, and obtain 
\begin{align*}
    \sum_{n=1}^N R^\textrm{u}_n=\sum_{n=1}^N\sum_{p=1}^N|U_{np}|^2+\sum_{n=1}^N\sum_{p=1}^M|U_{n\perp_p}|^2=&N,\\
    \sum_{n=1}^M R^\textrm{d}_n=\sum_{n=1}^M\sum_{p=1}^N|U_{\perp_np}|^2+\sum_{n=1}^M\sum_{p=1}^M|U_{\perp_n\perp_p}|^2=&M. 
\end{align*}
Thus, we have 
\begin{align*}
    N+M=&\sum_{n=1}^N\sum_{p=1}^N|U_{np}|^2+\sum_{n=1}^N\sum_{p=1}^M|U_{n\perp_p}|^2+\sum_{n=1}^M\sum_{p=1}^N|U_{\perp_np}|^2+\sum_{n=1}^M\sum_{p=1}^M|U_{\perp_n\perp_p}|^2\\
    =&\sum_{n=1}^N\sum_{p=1}^N|U_{np}|^2+\sum_{n=1}^M\sum_{p=1}^N|U_{\perp_np}|^2+M
\end{align*}
which has been obtained by using the fourth constraint $\sum_{p=1}^M C^\textrm{r}_p=M$ that appears in \eqref{solution:rows:columns}. This  immediately gives us
\begin{align}
    N=&\sum_{n=1}^N\sum_{p=1}^N|U_{np}|^2+\sum_{n=1}^M\sum_{p=1}^N|U_{\perp_np}|^2.
\end{align}
We now use the same argument as the one employed before to note that, for a large enough redshift $\chi$, we will inevitably have $|U_{np}|^2\leq C_{np}\chi^{-2}$ for appropriate constants $C_{np}$ independent of $\chi$. Thus,
\begin{align}
    N-\sum_{n=1}^M\sum_{p=1}^N|U_{\perp_np}|^2\leq&C\,\chi^{-2},
\end{align}
where we have introduced the constant $C:=\sum_{n=1}^N\sum_{p=1}^N C_{np}$. Crucially, $C$ is independent of $\chi$.

Finally, we have the constraint 
\begin{align*}
    \sum_{n=1}^M\sum_{p=1}^N|U_{\perp_np}|^2=M-\sum_{n=1}^M\sum_{p=1}^N|U_{\perp_n\perp_p}|^2\leq M.
\end{align*}
All of the results above imply the sequence of inequalities
\begin{align}
    N-M\leq N-\sum_{n=1}^M\sum_{p=1}^N|U_{\perp_np}|^2\leq C\chi^{-1},
\end{align}
which give us our final constraint
\begin{align}\label{unitarity:constraint}
    N-M\leq C\chi^{-1}.
\end{align}
Since \eqref{unitarity:constraint} applies for all $\chi\gg1$, we have that the matrix can be unitary for all redshift if and only if $N\leq M$. Thus, the smallest matrix $\boldsymbol{U}(\chi)$ that satisfies the constraints is a $2N\times2N$ matrix. 

We conclude that the proposal of an $N+N$-decomposition appears to solve the domain problem inherent in QOGRM and allows for employing the model for all values of the redshift. However, we note that by imposing the constraints \eqref{solution:rows:columns} we have effectively provided a necessary condition for a solution to exist: in fact, by enforcing  \eqref{solution:rows:columns} we are showing that the conditional solution of interest, i.e., the $N+N$-decomposition, does have the properties required and consequently is a necessary condition for unitarity. This comes at a price, which is the presence of at least $N$ environment modes, or perpendicular modes, that are not a priori known. In general, the knowledge of the exact nature of such modes might not be necessary, however, if it were the case that such knowledge was necessary, effort must be spent to obtain them. When the number of these perpendicular modes is low it is reasonable to expect that analytical expressions can be obtained. This might also occur if particular symmetries or other generic properties play a crucial role in the process. In all other cases, one might need to resort to numerical solutions. These issues are directly related to the arbitrariness available in the choice of the perpendicular modes. If it were possible, for example, to prove that such modes must have specific generic properties, given those of the $N$ modes of interest, this might aid in the quest of determining them when necessary. Understanding the degree of freedom in choosing the environment modes, together with studying their properties, is left to future work.

\section{Considerations and outlook}\label{considerations}
We have studied the validity of the QOGRM model first proposed in the literature \cite{Bruschi:Ralph:2014}, further developed in subsequent work \cite{Bruschi:Schell:2023,Alanis:Schell:Bruschi:2023}, to understand the effects of gravitational redshift on the quantum state of a photon. We have focused on the applicability of this model for all values of the redshift $\chi$, and we have found a simple family of scenarios where the transformation violates the assumption of unitarity. In particular, the aspect at the centre of the issue is the model's assumption that all unwanted degrees of freedom in the system, which is infinite dimensional, can be collected in one ``environment mode''. In the following, we provide considerations on the highlighted issue as well as on the proposed partial solution, and provide an outlook onto future work.

\subsection{Considerations on the validity of the model}
 The aspect to be considered is if the mode-mixing model QOGRM originally proposed to implement gravitational redshift in a quantum field theoretical scenario is correct in the first place  \cite{Bruschi:Ralph:2014}. That the original proposal is expected to have a regime of validity where it can be consistently applied can be deduced by the fact that all experimental evidence suggests that a photon propagating through a weakly curved spacetime, such as that surrounding the Earth, is always received as a single photon if sources of loss are ignored. In fact, modern experiments are employing single-photons to establish entangled links between distant nodes, a process that can be performed with reasonable success \cite{PhysRevLett.126.020503,li2024microsatellitebasedrealtimequantumkey,jaeken2025emulationsatelliteuplinkquantum,Ecker_2021,Picciariello_2025,Wu_2024,Krzic2023,Ma2012}. The received photons will, in general, be different from the ones that have been sent, although the number of photons is conserved. In the parlance of quantum optics, this process can be viewed as  a \textit{passive operation}, which is modelled as a mode-mixing operation \cite{Adesso:Ragy:2014}. Since no propagation in curved spacetime ($\chi=1$) means that the photon remains identical to itself, while a propagation along an infinitesimal distance is expected to induce a very small change in the state of the photon (equivalent to the small redshift regime $|\chi-1|\ll1$), we conclude that gravitational redshift can be modelled as a continuous map around the identity operator $\boldsymbol{U}(\chi=1)=\mathds{1}$. The conclusion one reaches following this argument is that QOGRM can be correct at least in the regime of very small redshift. Note that here we have assumed that any potential effect of particle creation is nonexistent or negligible, since particle creation would imply that the transformation is \emph{active}, and the matrix $\boldsymbol{U}(\chi)$ symplectic but not unitary. That time-dependent frequency-shifts might lead to particle creation has been discussed in the context of observers that do not share the same vacuum state \cite{Alsing:Milonni:2004}. While such scenarios are not considered here, because we opt to focus on setups where observers share the same vacuum, it is an open question to understand if and when gravitational redshift can be associated to not only mode mixing, but also particle creation.

We then note that gravitational redshift is an effect that has been first derived in the context of classical (in the sense of quantum mechanics) physics \cite{Einstein:1908,Wald:1995}, and it is within this approximation---where quantum aspects can be ignored---that the effects have been measured \cite{bothwellResolvingGravitationalRedshift2022,PhysRevLett.45.2081,zhengLabbasedTestGravitational2023, mullerPrecisionMeasurementGravitational2010}. Nevertheless, photons are intrinsically relativistic and quantum systems, and therefore it is natural to seek an extension of the effect as put forward in the seminal work on this topic \cite{Bruschi:Ralph:2014}. While the original proposal has been expanded and developed in subsequent work \cite{molaei2024photongravitycouplingschwarzschildspacetime,PhysRevD.107.125010,PhysRevD.105.084016,mieling2023gupta}, the back-reaction of the field excitation on the gravitational field, and therefore on the quantum field itself, has not been considered. This is standard in works of quantum field theory in curved spacetime, where the fields of interest are considered to be \textit{probe fields}, i.e., they are assumed not to carry enough energy to bend spacetime comparably to other distributions of mass and energy \cite{Birrell:Davies:1982}. Therefore, QOGRM can be considered a stepping stone towards the understanding of the nature of the gravitational redshift in a fully relativistic and quantum theory at least at low energies, where the interaction between the photon and the gravitational field must be taken into account.

Another important observation concerns the role of the mode structure of the photons. Our work indicates clearly that the smallness of the redshift, i.e., how close $|\chi-1|$ is to $0$, is not enough to determine the fact that a regime under considerations allows to write $\boldsymbol{U}(\chi=1+\epsilon)\approx\mathds{1}+\boldsymbol{U}^{(n)}\epsilon^n$ to lowest order. 
To understand this aspect we note that perturbation theory can be applied only if the resulting expressions are such that the product of the coefficients in front of the perturbative parameter with the corresponding order parameter itself, remain small. Concretely, $u^{(n)}\epsilon^n\ll1$ for all $n$ and $u^{(n+1)}\epsilon^{n+1}\ll u^{(n)}\epsilon^n$. In the case under consideration, the coefficients $u_{nm}^{(1)}$ of $\boldsymbol{U}^{(1)}$ are expected to satisfy $u_{nm}^{(1)}\epsilon\ll1$ for all $u_{nm}^{(1)}\neq0$. However, it can well be that, given a fixed $\epsilon$, $u_{nm}^{(1)}\epsilon=\mathcal{O}(1)$ for certain mode configurations since these coefficients do depend on the mode structure. We provide an illustrative example to better clarify this aspect in the example below. 

\begin{tcolorbox}[colback=teal!3!white,colframe=teal!85!black,title=Example: Deformation of Gaussian wavepackets with complex phase]
Let us assume that $F(\omega)=(\sqrt{2\pi}\sigma)^{-1/2} e^{-(\omega-\omega_0)^2/(4\sigma^2)}e^{-i\omega\varphi}$ is a normalized Gaussian mode function. Then, we compute $\Delta(\chi)\equiv\langle F|F'\rangle$, and we obtain
\begin{align*}
    \Delta(\chi)\equiv\frac{\sqrt{2}\chi}{\sqrt{\chi^4+1}}e^{-\frac{1}{4}\frac{(\chi^2-1)^2}{\chi^4+1}\frac{\omega_0^2}{\sigma^2}}e^{-\frac{(\chi^2-1)^2}{\chi^4+1}\sigma^2\varphi^2}e^{-i\frac{\chi^4-1}{\chi^4+1}\varphi\omega_0}.
\end{align*}
It is immediate to verify that $\Delta(\chi=1+\epsilon)\approx1-2i\varphi\omega_0\epsilon$ and $|\langle F'|F_\perp\rangle|=|\Delta(\chi=1+\epsilon)|\approx1-\frac{1}{4}((\frac{\omega_0}{\sigma})^2+4\sigma^2\varphi^2)\epsilon^2$ to lowest nontrivial order, as already observed in the literature  \cite{Bruschi:Ralph:2014}. Here, we can immediately identify the first element of the $\boldsymbol{U}(\chi)$ matrix with $\Delta(\chi=1+\epsilon)$, which confirms that the first-order correction is purely imaginary as predicted above. Note that $\varphi\omega_0\epsilon\ll1$ as well as $\alpha^{(2)}\epsilon^2\ll1$ for the perturbative expansion to be valid, where $\alpha^{(2)}$ is the coefficient of the second order term. It is immediate to see that one cannot naively evaluate $|\Delta(\chi=1+\epsilon)|$ in its perturbative form for arbitrary values of $\phi$. In fact, while the full analytical expression guarantees that $|\Delta(\chi=1+\epsilon)|\geq0$ for all $\varphi$, its total expression to second order in $\epsilon$ can become negative for sufficiently large values~of~$\varphi$.
\end{tcolorbox}

As mentioned above, if we were to take the first order contribution of $\Delta(\chi=1-\epsilon)\approx1+2i\varphi\omega_0\epsilon$ for all values of $\varphi$ it might seem that this term can grow linearly indefinitely as a function of $\varphi$. However, it is clear from the full expression $\Delta(\chi)$ that such unbounded linear growth cannot occur. This is but one simple example that highlights the subtleties of perturbative expansions.

We now move on to discussing the intriguing interplay of deformation and phase shift. The interplay of \textit{genuine deformation} and \textit{genuine redshift} has already been introduced and studied in the literature \cite{Bruschi:Chatzinotas:2021}. The former is defined as the maximum value of $|\Delta(\chi)|$ obtained by taking the transformed mode $F'$ and optimizing over \textit{rigid shifts} of the frequency spectrum as a whole. The amount of translation in the frequency domain that is necessary to optimize the overlap was defined as the genuine redshift. In this work we have found that the perturbative expansion \eqref{delta:perturbative:expansion} of $\Delta(1+\epsilon)$ tells us that $|\Delta(\chi)|$ is always obtained at second order in $\epsilon$, in agreement with preliminary results in the literature \cite{Bruschi:Chatzinotas:2021}. More work is necessary to understand mode-overlap optimization in the slightly more general context presented here.

The quest for characterizing deformation and shift of the wavepacket brings us to make the following crucial observations: first, when $F$ is real its phase $\theta$ vanishes and it immediately follows that $\kappa=-\int \textrm{d}x\,x\,\dot{\theta}(x)\, \rho^2(x)=0$ as well as $\mu^2=\int \textrm{d}x\,x^2\dot{\rho}^2(x)$, and therefore $F=1+\mathcal{O}(\epsilon^2)$. The role of the phase $\theta$ was already discussed in the literature \cite{Bruschi:Chatzinotas:2021}. Second, there is an optimal value $\kappa^\text{opt}$ of $\kappa$ for which $\Delta_\epsilon=e^{2i\kappa^\text{opt}(1-\frac{\epsilon}{2})\epsilon}$ and $|\Delta_\epsilon|=1$ to order $\epsilon^2$. In this case, we see that there is no deformation to the first two nontrivial perturbative orders, however, gravitational redshift acts as an overall phase shift in the mode as long as $\kappa^\text{opt}\neq0$. That the overall effect vanishes completely only occurs when $\kappa^\text{opt}$ also vanishes, which requires $\mu^2=\int \textrm{d}x\,x^2\,\dot{\rho}^2(x)=1/4$. Most importantly, the value of $\kappa^\text{opt}$ depends solely on the properties of the modulus $\rho=|F|^2$ and is independent on the redshift. With these intriguing preliminary considerations in mind, we leave it to further work to study the resilience of optical modes to being transformed by the effects of gravity, as well as to engineer frequency profiles that can be used to either maximize or minimize the effects according to the desired outcome.

\subsection{Applicability within and beyond the perturbative regime}
The quantum optical model QOGRM studied here, including the $N+M$-decomposition of the unitary matrices proposed to extend the domain of validity of the original model, can be validated experimentally only for small values of the gravitational redshift. Interestingly, the magnitude of the effects does not only depend on the redshift itself, which is a dimensionless quantity, but it also depends on the parameters of the mode function as well. It can come as a surprise, at a superficial glance, that the magnitude of a dimensionless quantity such as the redshift parameter is not sufficient to determine the scenarios within which its contribution is de-facto small, in particular within a framework where the effect should be determined by a background configuration (i.e., the spacetime) that is external to, and independent from, the system itself. Nevertheless, it should be clear from all of our considerations in this work that, since the transformation of interest, as an effect, moves and deforms the wavepackets of the photons in frequency space, it follows that the magnitude of the overlap of initial and transformed wavepacket will in general depend on the wavepacket itself. While we have already stressed this point multiple times before, we add here another dimension to this aspect. General relativity and quantum mechanics are, as far as our current understanding goes, completely independent theories.\footnote{We do not enter here the debate of determining a theory of quantum gravity \cite{Kiefer:2012}, since this is not relevant to the energy scales where the effects discussed in this work could be tested in the near future.} Quantum field theory in curved spacetime is perhaps our best approach to study the dynamics of quantum systems when they are placed in a background dynamical spacetime \cite{Birrell:Davies:1982}. Excitations of quantum fields are treated, as already mentioned, as small perturbations that do not back-react on the configuration of spacetime. Within this context, gravitational redshift is a classical effect that is predicted by general relativity. It remains to some degree unclear if the redshift is effectively experienced by the photon, or if it originates from the mismatch of frequencies as measured locally by two users placed at different locations in a curved background \cite{Okun:Selivanov:2000,Okun:2000}. 
In this sense, it is by no means self evident that such effect should be visible in the kinematics of propagating quantum photons. Even more strikingly, it is also not self evident that quantum properties of the propagating systems should be sensitive to the gravitational redshift in the first place. Therefore, we find it intriguing that the quantum state of a photon deforms differently as a function of its defining properties, in particular quantum coherence (i.e., a nonzero phase $\theta$ in our work). An important observation is the following: if this effect cannot be recovered by studying the interaction of propagating photons in curved spacetime with gravity, we would be in principle able to discriminate between ``redshift as a local frequency mismatch'' and ``redshift due to propagation''. We can thus say that, while we have found a necessary condition to solve the problem by including at least one auxiliary mode $F_\perp$ for each mode $F$ of interest, not only it is remains unclear how to determine the space of solutions $\{F_{\perp_n}\}$, but it is to date not possible to conclude whether this model would pass the test of experimental validation if it were possible to measure the effects of very large gravitational redshift, such as in the case of photons emitted close to the event horizon of a black hole on realistic photons \cite{Misner:Thorne:1973}, since it would be necessary to know the frequency profile at the point of emission in order to compute the overlaps $\Delta_{nm}$. We leave these questions, including how this phenomenon occurs, what mechanism enables it, and what occurs beyond weak gravity (or small redshift) to future work.

\subsection{Impact on current efforts in quantum technologies}
We have argued in this work that it is possible to extend the validity of QOGRM to the full redshift domain.
A natural question to ask is if this has a measurable effect on existing or future technologies, in particular in view of the planned deployment in space of parts of a global quantum network \cite{Simon_2017,Liao:Yong:2017,Gundogan:Sidhu:2021}. 
There is an increasing body of work dedicated to answer this question. In the past years it has been shown that the physical process discussed here can in principle affect, for example, quantum key distribution and quantum communications \cite{ Liao:Cai:2017,Yin:Ren:2017,Zhang:Xu:2018,Bedington_2017,Sidhu2021,Pirandola:2021,Cao:Yin:2023}. In such cases, discrimination of states in terms of their spectral properties is crucial for the success, and therefore any deviation from perfect distinguishability or indistinguishability might lead to the degradation of the channel. Additionally, it has also been argued that a change in the quantum state can be employed in quantum metrological schemes to improve the measurement sensitivity of relevant physical parameters, such as distances, planetary masses, and rotation frequencies \cite{Kohlrus:Bruschi:2017,Kohlrus:Bruschi:2019}. Furthermore, the predicted state changes might accumulate during the distributed processing of complex quantum information within complex 
scenarios revolving around quantum information processing for the quantum internet ~\cite{kimble:2008,gibney:2016}, thereby leading to increased error rates that need to be taken into account and corrected for.

While existing sources of noise and loss are the predominant effects to be taken into account in the design of future quantum communication technologies \cite{Sharma:Banerjee:2019,Chai:Cao:2019,Hausler:Orsucci:2024,MarulandaAcosta:Dequal:2024,Ajith:Veni:2025}, effects of gravitational redshift can potentially become measurable within dedicated experiments since the magnitude of the effects need not be insignificant\cite{Bruschi:Ralph:2014}. Given the constant improvement of the performance of photonic sources and capability of measurement devices, it is reasonable to believe that it is only a matter of time before gravitational redshift deformations of quantum states of photons will be measurable.

\subsection{Outlook}
We conclude by providing a brief and not exhaustive outlook for this work.  
In order to better understand physical phenomena relevant to this work, it is desirable to predict gravitational redshift from a more fundamental perspective of relativistic and quantum physics. One potential avenue is that of quantizing linearized gravity coupled to matter or fields, that is, an approach where the background gravitational field is a perturbation of flat Minkowski spacetime and the fields are both quantized \cite{guptaQuantizationEinsteinGravitational1952}. The resulting Hamiltonian contains an interaction term between the stress energy tensor of a (quantum) field. This approach has been key in recent times to studies of the interaction of gravitons with matter in the context of the interplay between the effects of quantized gravity and matter dynamics \cite{PhysRevLett.119.240401,marshmanLocalityEntanglementTabletop2020,RevModPhys.97.015003}. An initial effort in this avenue has recently produced interesting preliminary results   \cite{Lapponi:Ferreri:2025}. Progress in this direction can aid our understanding of the interplay of general relativity and quantum mechanics in the low energy and weak curvature regime.

It is also important to briefly outline potential alternative avenues for testing the effects predicted here.
One such avenue of study is an analogue model where the kinematics of a system that is controllable in a laboratory mimic those of interest \cite{Barcelo:Liberati:2005}. Regarding the case of interest here, a simple scenario could include the electromagnetic field propagating within a nonlinear medium that reproduces, at least qualitatively, the effects. Nonlinear media are known to provide an ideal test-bed for many applications, including the aforementioned studies of analogue gravity \cite{PhysRevD.65.064027}. Pursuing this avenue would have the added advantage that, if a suitable toy-model can be devised whose predictions match those of this work at least in the weak-gravitational regime, such predictions could also be measured in an Earth-based experiment. This would not only reduce dramatically the costs of developing a first-of-its-kind experiment using conventional satellites, but it would also provide a proof-of-principle demonstration to be used in the design of more advanced ground-to-satellite or satellite-to-satellite experiments.
Another avenue for testing the predictions of this work has been opened by the development of the  CubeSat and Nanosat technology \cite{Saeed:Elzanaty:2020,Kopacz_2020,Araniti:Iera:2022}. 
These satellites are small and inexpensive as compared to their conventional counterparts, and they enjoy a customizable plug-and-play approach to satellite and payload design, thus enabling affordable devices with reasonable lifespans. Such platforms are already used for different purposes within academia and industry, including quantum science in space \cite{Tang:Chandrasekara:2016,Bedington:Bai:2016,Oi:Ling:2017}, and it can be envisioned that a single photon source with the desired specifications could be loaded on a CubeSat to be used for proof-of-principle experiments.

\section{Conclusions}\label{sec conclusions}
In this work we have studied the domain of validity of a recently proposed quantum-optical model of gravitational redshift, called QOGRM, that can been implemented as a multimode-mixing gate on photonic quantum states \cite{Bruschi:Ralph:2014}. We have found that the domain of validity of the model, in its current form, does not extend to all values of the redshift but is limited to small redshift conditional to the defining parameters of the photon. Such parameters include those determining the mode structure of the photons, the distance between users exchanging photons, and the strength of the local gravitational field. We identified the nature of the problem, which laid in the breakdown of unitarity of the transformation for large redshift due to a non-injective map between the modes of interest and a single artificial environment mode, and we obtained a necessary condition to solve the problem based on extending the unitary transformation to include a sufficient number of auxiliary modes to restore unitarity. This condition comes at the price of determining the properties, when such properties are required, of the auxiliary modes, whose number scales only linearly with the number of input modes of interest. Tackling this  issue can be done analytically for small numbers of modes or using numerical methods in the case of large numbers of modes. 

We concluded our work by providing a discussion on the implications of our results for the study of gravitational redshift as a quantum phenomenon in particular, for the understanding of relativistic quantum physics in the low energy and weak gravity regime more broadly, as well as on the impact on planned space-based quantum technologies.

\section*{Acknowledgments}
We thank Alessio Lapponi for useful comments. DEB would like to acknowledge constructive feedback received on his presentation of the findings reported in this work at the \hyperlink{https://indico.math.cnrs.fr/event/11461/}{Avenues of Quantum Field Theory in Curved Spacetime 2025} conference, as well as comments from anonymous referees that have greatly improved of this work.

\section*{Declarations}



\subsection*{Funding}
A.F. and D.E.B. acknowledge support from the joint project No. 13N15685 ``German Quantum Computer based on Superconducting Qubits (GeQCoS)'' sponsored by the German Federal Ministry of Education and Research (BMBF) under the \href{https://www.quantentechnologien.de/fileadmin/public/Redaktion/Dokumente/PDF/Publikationen/Federal-Government-Framework-Programme-Quantum-technologies-2018-bf-C1.pdf}{framework programme
``Quantum technologies -- from basic research to the market''}. D.E.B. also acknowledges support from the German Federal Ministry of Education and Research via the \href{https://www.quantentechnologien.de/fileadmin/public/Redaktion/Dokumente/PDF/Publikationen/Federal-Government-Framework-Programme-Quantum-technologies-2018-bf-C1.pdf}{framework programme
``Quantum technologies -- from basic research to the market''} under contract number 13N16210 ``SPINNING''. 

\subsection*{Conflict of interest/Competing interests}
Not applicable.

\subsection*{Ethics approval and consent to participate}
Not applicable.

\subsection*{Consent for publication}
DEB and AF have requested and received consent for publication as required by the statute and bylaws of Forschungszentrum J\"ulich GmbH.

\subsection*{Data availability}
Not applicable. 

\subsection*{Materials availability}
Not applicable.

\subsection*{Code availability}
Not applicable.

\subsection*{Author contribution}
DEB and AWS conceived the original idea. NL and LAAR worked on the original idea. NL identified the existence of a limit to the validity of the original model. DEB, AWS and AF proposed an analysis of the problem. All authors contributed to writing the manuscript.

\bibliography{References}

\end{document}